\documentclass[aps,prb,showpacs,superscriptaddress,twocolumn]{revtex4}
\usepackage{graphicx}
\usepackage{amsmath}
\usepackage{bm}
\usepackage{xcolor}

\begin{document}

\title{The focusing effect of electron flow and negative refraction in three-dimensional topological insulators}

\author{Kai-Tong Wang}
\affiliation{School of Physics, Beijing Institute of Technology, Beijing 100081, China}

\author{Yanxia Xing}
\email[]{xingyanxia@bit.edu.cn}
\affiliation{Beijing Key Laboratory of Nanophotonics and Ultrafine Optoelectronic Systems, School of Physics, Beijing Institute of Technology, Beijing 100081, China}

\author{King Tai Cheung}
\affiliation{Department of Physics and the Center of Theoretical and Computational Physics, The University of Hong Kong, Pokfulam Road, Hong Kong, China}

\author{Jian Wang}
\affiliation{Department of Physics and the Center of Theoretical and Computational Physics, The University of Hong Kong, Pokfulam Road, Hong Kong, China}

\author{Hui Pan}
\affiliation{College of Physics, Beihang University, Beijing 100191, China}

\author{Hong-Kang Zhao}
\email[]{zhaohonk@bit.edu.cn}
\affiliation{School of Physics, Beijing Institute of Technology, Beijing 100081, China}

\begin{abstract}
We studied the focusing effect of electron flow induced by a single p-n junction (PNJ) in three-dimensional topological insulator. It is found that the electrons flowing from the n region can be focused at the symmetric position in the p region, acting as a perfect Veselago lens, regardless whether the incident energy is within or beyond the bulk energy gap. In the former case, the focusing effect occurs only in the surfaces.
While in the latter case, the focusing effect occurs beyond the surfaces.
These results show that the focusing effect of electron flow is a general phenomenon.
It means the negative refraction may arise in all materials that are described by the massive or massless Dirac equation of 2D or beyond 2D system. Furthermore, we also find the focusing effect is robust in resisting the moderate random disorders. Finally, in the presence of a weak perpendicular magnetic field $B_z$, the focusing effect remains well except that the position of the focal point is deflected by the transverse Lorentz force. Due to the finite size effect, the position of focal point oscillates periodically with a period of $\Delta B\approx\frac{h/e}{W_xW_y}$.
\end{abstract}

   % Conductivity phenomena in semiconductors and insulators
%75.50.Pp	% Magnetic semiconductors
%72.20.Ee,   % mobility edges, hopping transport
%73.43.Qt	% Magnetoresistance
%73.63.-b    % Electronic transport in nanoscale materials and structures
%73.40.-c	% Electronic transport in interface structures

\maketitle
\section{INTRODUCTION}
In 1960s, Veselago theoretically predicted the existence of negative refractive index material, i.e., left-handed material.\cite{Veselago,Ward2005} After about 30 years, the first artificial left-handed material was experimentally verified.\cite{Science292.77-79.Shelby.2001}
%Since then, NRIM has attracted increasingly great interest in the field of artificial metamaterials.
%So,it has the potential for making devices superior to conventional ones. %Considering the refractive index $n=\sqrt{\epsilon\mu}$ is real, both $\epsilon$ and $\mu$ are negative, which leads to the reversal of phase and group velocity
In general, electromagnetic negative refraction can only be realized in artificial constructed metamaterials because of the negative $\epsilon$ and $\mu$,\cite{Science305.788-792Smith2004} where $\epsilon$ and $\mu$ are the electric permittivity and the magnetic permeability, respectively. However, concerning the wave vector and group velocity, the  optical rays and the electron flow (electron's de Broglie wave) are similar, so the negative refraction would be achieved in real massless Dirac fermion materials, such as graphene. In a real material, the negative refraction is directly related to perfect Veselago lens\cite{Phys.Rev.Lett.85.3966-3969Pendry2000} and Klein paradox.\cite{Phys.Rev.A79.Guney2009} %a counterintuitive phenomenon in relativistic quantum mechanics.\cite{Klein}

The existence of negative refraction in massless Dirac material is natural. The electrons and holes in massless Dirac material are conjugately linked and interconnected, the chiralities (or dispersions) in conductance band and valence band are opposite. Then, the potential barrier induced by p-n junction (PNJ) is highly transparent for the charge carriers,\cite{Katsnelson2006a} As a result, the electron flow would be negatively refracted and symmetrically focused by the straight interface of PNJ in the linear dispersion region.\cite{Cheianov2007a} Beyond the linear region,
the statement on Dirac fermion fall through, however, the focusing effect exists still.\cite{Xing2010} It means the negative refraction is not limited to the two dimensional massless Dirac materials. In fact, as shown in Fig.1, when electrons with momentum $(k_x,k_\parallel)$ and velocity $(v_x,v_\parallel)$ penetrate through PNJ and become holes with momentum $(-k_x,k_\parallel)$, due to the opposite dispersion for electrons and holes, the velocity of holes becomes $(v_x,-v_\parallel)$, then the negative refraction is formed. As a result, the electron flow is focused by the straight interface induced by PNJ. Here, '$\parallel$' denotes the direction along $y$ for two dimensional system or $y$-$z$ plane for three dimensional system. So, there are two essential conditions to the focusing effect of electron flow. One is the opposite dispersions in conductance band and valence band, the other is the nearly transparent PNJ.
In principle, besides massless Dirac Fermions\cite{PhysicalReviewB90.035138.Fleury.2014,Pendry2012}, all gapless semi-metal and topological materials\cite{PhysicalReviewB92.041408.Zhao.2015,PhysicalReviewB94.075137.Sessi.2016} described by quadratic massive Dirac equation in two dimension (2D) or beyond 2D, such as the 3D topological insulator (TI),
ought to have the same effect. Considering the helical resolved characters of the TI materials, the focusing effect in TI can have great potential in the applications of helicity-based electron optics\cite{PhysicalReviewB92.041408.Zhao.2015}.

\begin{figure}
\includegraphics[width=8.7cm, clip=]{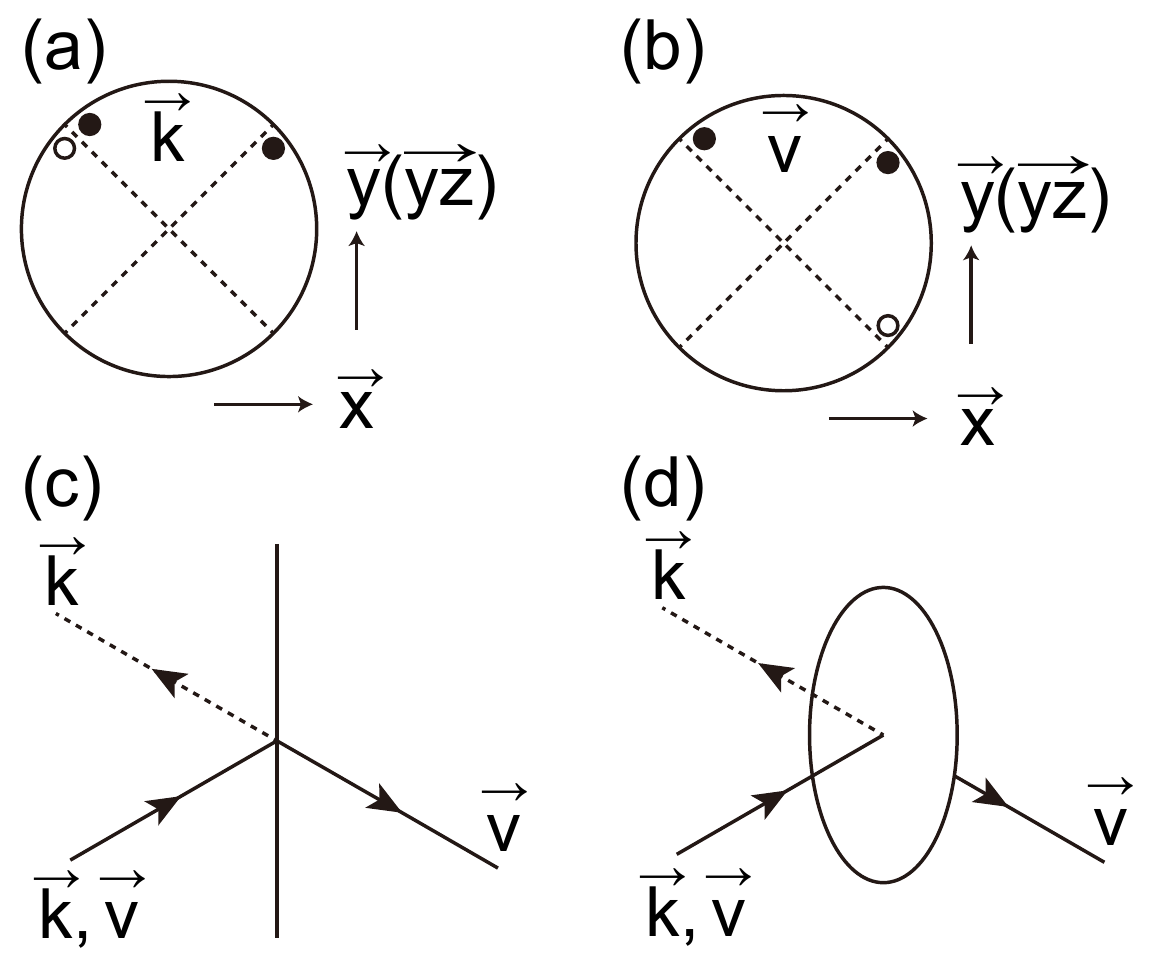}
\caption{ (Color online) An electron with momentum $(k_x,k_\parallel)$ and velocity $(v_x,v_\parallel)$ is injected from the left $n$ region and scattered by the one-dimensional PNJ [panel (c)] or two-dimensional PNJ [panel (d)] to the right $p$ region as a hole (hollow black dots) with momentum $(-k_x,k_\parallel)$ and velocity $(v_x,-v_\parallel)$ due to the opposite dispersion for electrons and holes, which leads to the focusing effect of electron flow. The corresponding momentum and velocity of electrons (solid circles) and hole (open circles) is shown in the panel (a) and panel (b), respectively.}
\end{figure}

In the past years, due to the extraordinary band structure and huge potential in making future devices, TI have attracted great attentions\cite{Rev.Mod.Phys.82.3045-3067.Hasan.2010,Qi2011,Science301.1348-1351Murakami2003,Bernevig2006,Konig2007,Chen2009,Zhang2009,Science329.61-64.Yu.2010,Science340.167-170.Chang.2013,Zhang2010,Shan2010}
in condensed matter physics.
%In the band gap of TI, the bulk electron behaves as an ordinary insulator because of the separated conduction and valence bands. Near the boundary, TI is conducting since a couple of helical boundary states reside stably within the bulk energy gap.
For a 3D TI such as $(Bi,Sb)_2Te_3$,\cite{Chen2009,Zhang2009} the electrons on conducting surface are massless Dirac Fermions depicted as a single Dirac cone.
On the other hand, when the fermi energy is beyond the energy band gap, TI is similar as a conventional semiconductors because of the separated bands (conductance band and valence band).
%By controlling the gate voltage or doping technique, the PNJ can be fabricated in the 3D TI.%\cite{XXX}
Then, what about the focusing effect of electron flow in 3D TI considering these different states of matter?
can we find the focusing effect in deep conductance/valence band?

To answer these questions, we've constructed the PNJ in an infinite 3D TI ribbon as shown in Fig.2(a).
%The 3D TI is depicted by the tight binding Hamiltonian using band parameters of $(Bi,Sb)_2Te_3$ compound.
With the aid of non-equilibrium Green's function, we study the local conductance response to the non-equilibrium electron injection in 3D TI with single PNJ. It is found that when the incident energy is in the bulk gap, the transport processes are dominated by the surface states, and the focusing effect then arises only on the surfaces.
As shown in Fig.2(a), on each side surface, electrons flow injected (the blue points) from n region (blue region) can be focused in the p region (red region) in the symmetric position (the red points). It is not strange since the surface states of 3DTI satisfy the 2D massless Dirac equation. When the incident energy is beyond the bulk energy gap, TI resembles the conventional semiconductors. However, because of the conjugated interconnection between the conductance band and valence band, the focusing effects in the bulk is even better. In this case, the electron flow incident from any site with the position of $(-x,y,z)$ in the n region would be focused at position $(x,y,z)$ in the p region.
Although supported by both surface and bulk states, the focusing effect can not be observed when these two type of states are mixed (near the energy band edges), because of the different dispersion for surface states and bulk states.
Furthermore, we have also studied the influence of random scattering and the weak external magnetic field $B_z$ on the focusing effect. It is found that the focusing effect is immune to random disorders. In the presence of weak perpendicular magnetic field $B_z$, the focus point is deviated by the lateral Lorentz force, however, the focusing effect retains well. Owing to the finite size of the scattering region, with the increasing $B_z$, the position of the focus oscillates periodically with the period of $\Delta B\approx\frac{h/e}{S}$, where $S$ is the area of central p region.

The paper is organized as follows. In Sec. \uppercase\expandafter{\romannumeral2}, from the low energy effective model, we present the system Hamiltonian in real space using tight binding technique. Then, both the partial local density and the local conductance describing the local response to the non-equilibrium source, i.e., the incident electron flow, are derived.
Sec. \uppercase\expandafter{\romannumeral3} is the numerical results and some
discussions. Finally, a summary of our work is presented in Sec.
\uppercase\expandafter{\romannumeral4}.

\section{MODEL AND FORMALISM}

Through $k\cdot p$ perturbation, the low energy effective Hamiltonian of 3D TI can be expanded in the Hilbert space composed with four low-lying states at $\Gamma$ point, i.e., $|P1_{z^+,\uparrow}\rangle$, $|P2_{z^-,\uparrow}\rangle$, $|P1_{z^+,\downarrow}\rangle$ and $|P2_{z^-,\downarrow}\rangle$.
Correspondingly, the Hamiltonian of infinite 3D TI is written in the following form: \cite{Zhang2009,Liu2010,Qi2011}
\begin{equation}
  H_{0}(k)=\epsilon_{k}+M_{k}\sigma_{0}\tau_{z}+A_{\bot}k_{z}\sigma_{z}\tau_{x}+A_{\|} (k_x\sigma_x+k_y\sigma_y)\tau_{x}\\%\nonumber
\end{equation}
where, $\sigma_\alpha$ and $\tau_\alpha$ represent the real spin ($\uparrow$ and $\downarrow$) and pseudo-spin (signing the orbital $|P1_{z^+}\rangle$ and $|P2_{z^-}\rangle$) with $\alpha=x,y,z$. $\sigma_{0}$ and $\tau_{0}$ are $2\times 2$ unitary matrix. $\epsilon_{k}=C_{0}+C_{\bot}k_{z}^2+C_{\|}(k_x^2+k_y^2)$, $M_{k}=-D_{0}+D_{\bot}k_z^2+D_{\|}(k_x^2+k_y^2)$.
%$A_x=A_y=A_{\parallel}$, $A_z=A_{\perp}$, $D_x=D_y=D_{\parallel}$, $D_z=D_{\perp}$.
Here, we set $\epsilon_k=0$ since it shifts the Dirac point and doesn't change the topological structure of the Hamiltonian. To investigate the spacial focusing effect, the Hamiltonian expressed in real space is needed. Replacing $k_{x,y,z}$ by $-i\nabla_{x,y,z}$, we get 3D effective tight-binding Hamiltonian in a square lattice, as follows:\cite{Zhang2014a}
\begin{equation}
H_0 = \sum_{\mathbf{i}}d_{\mathbf{i}}^{\dag}H_{\mathbf{i}}d_{\mathbf{i}}
+\sum_{\mathbf{i},\alpha}d_{\mathbf{i}}^{\dag}H_{\alpha} d_{\mathbf{i}+\mathbf{a}_{\alpha}}+H.c. \label{H0}
\end{equation}
with
\begin{equation}
\begin{split}
  H_{\mathbf{i}} &=\epsilon_{\mathbf{i}}\sigma_{0}\tau_{0}+(D_{0}+2\sum_{\alpha}\frac{D_{\alpha}}{a^2})\sigma_{0}\tau_{z} \\
  H_{\alpha} &=\left[-\frac{D_{\alpha}}{a^2}\sigma_{0}\tau_{z}-i \frac{A_{\alpha}}{2a}\sigma_{\alpha}\tau_{x}\right]e^{i\phi_{\mathbf{i},\mathbf{i}+a_{\alpha}}}\nonumber
  \end{split}
\end{equation}
where, $\alpha=x,y,z$, $d_{\mathbf{i}}=[d_{\mathbf{i},1_z^+,\uparrow},d_{\mathbf{i},2_z^-,\uparrow},d_{\mathbf{i},1_z^+,\downarrow},d_{\mathbf{i},2_z^-,\downarrow} ]$ denotes the four low-lying states at $\Gamma$ point, $\epsilon_{\mathbf{i}}$ is the onsite energy at each lattice site. Here, $\mathbf{i}=[\mathbf{i}_x,\mathbf{i}_y,\mathbf{i}_z]$ is used to indicate the discrete sites of the square lattice with lattice constant $a$. Considering the perpendicular magnetic field $B_{z}$, the extra phase $\phi_{\mathbf{i},\mathbf{i}+a_{\alpha}}=\frac{e}{\hbar}\int_{\mathbf{i}}^{\mathbf{i}+a_{\alpha}}\mathbf{A}\cdot \mathbf{dl}$ is induced\cite{Chen2012,Zhang2014} by the magnetic vector potential $\mathbf{A}$. In the Coulomb gauge, the vector potential is set as $\mathbf{A}=[-By,0,0]$ and the magnetic flux at each lattice is then $\Phi_0=B_za^2$. For an infinite nanoribbon shown in Fig.2(a), $y$ and $z$ are finite and $x\in[-\infty,\infty]$.
The incident electrons are free in the $x$-direction if PNJ is absent.

\begin{figure}
\includegraphics[width=7cm,clip=]{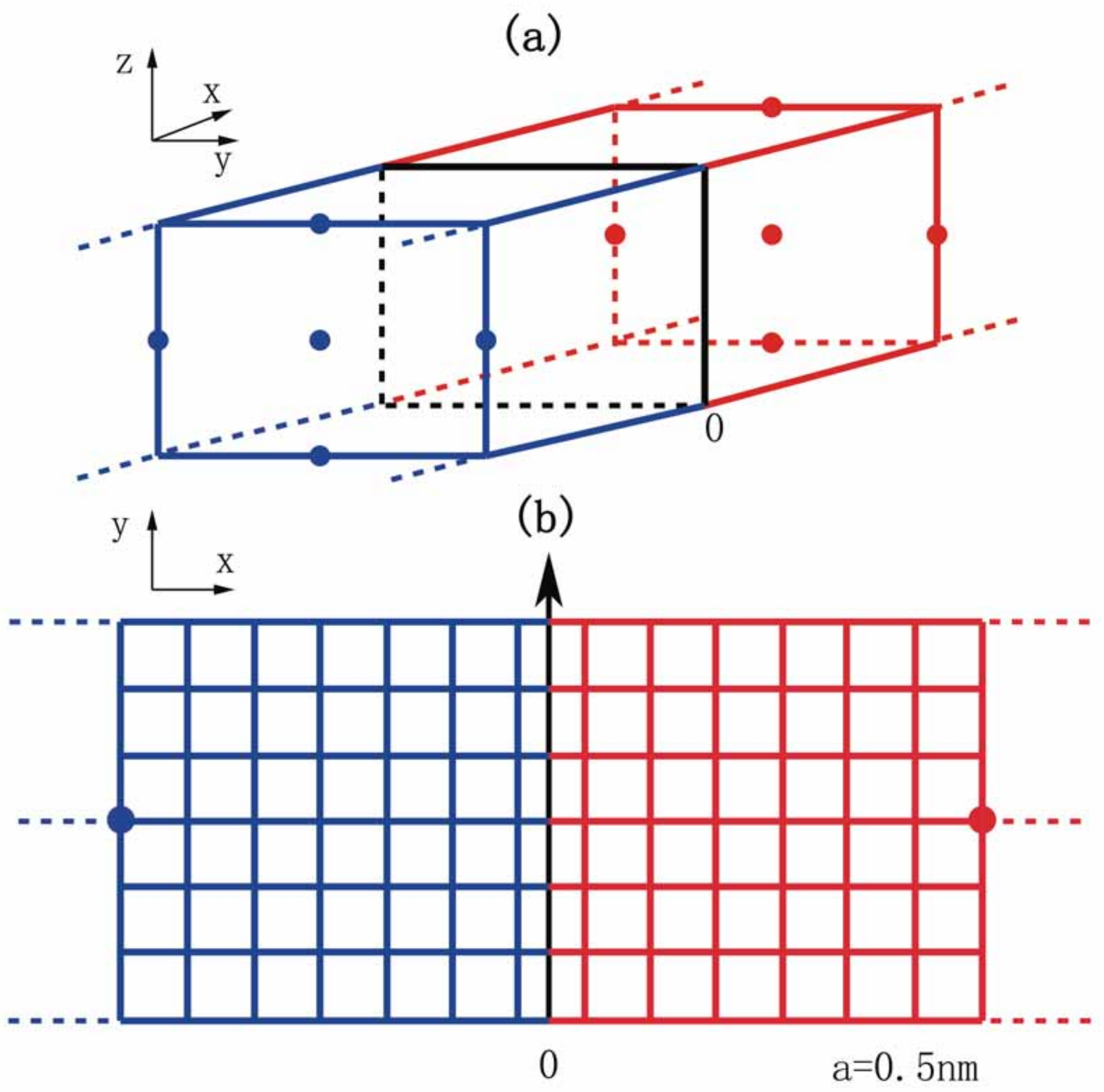}
\caption{ (Color online) Panel (a): schematic diagram of two-terminal open system composed by semi-infinite n region (green region) and semi-infinite p region (orange region) with a sharp PNJ located at $x=0$ (the black square). For the bulk states, the incident electron flow is (the Green point) scattered by PNJ and symmetrically focused in the P region (the yellow point). For the surface states, the focusing effect occurs only in the surfaces, the injected electron flow and the focused electron flow are signed by the blue points and red points in the surfaces. Panel (b): the focusing effect in the side surface.}
\end{figure}

In the presence of the sharp PNJ that is induced by a step potential $U(x)=E_0[\theta(x)-\theta(-x)]$, the system is composed with semi-infinite electron-like n region ($x<0$, the blue region in Fig.2(a) where $\epsilon_{\mathbf{i}}=-E_0$) and semi-infinite hole-like p region ($x>0$, the red region in Fig.2(a) where $\epsilon_{\mathbf{i}}=E_0$). Because the semi-infinite p region and n region are ideally periodical, the incident electrons can only scattered by the straight interface [the black interface in the Fig.2(a)], i.e., the sharp PNJ located at $x=0$. Here, electrons are locally injected through source terminal $H_s$ and detected through drain terminal $H_d$. The total Hamiltonian including the source and detection terminals is then expressed as
\begin{equation}
  H=H_0+H_{s}+H_{d}+H_{T,s}+H_{T,d}\label{H1}
\end{equation}
with
\begin{equation}
\begin{split}
&H_{s/d}=\sum_{k}\epsilon_{s/d,k}c_{s/d,k}^\dagger c_{s/d,k}\\
&H_{T,s/d}=\sum_{k,\mathbf{i}_s}t_{s/d}[d_{\mathbf{i}_{s/d}}^\dag c_{s/d,k}+H.c.]\nonumber
\end{split}
\end{equation}
where $H_0$ denotes the infinite nanoribbon with sharp PNJ. $H_{s/d}$ denotes the source or detecting terminal used to inject or detect electron flow. Phenomenologically, $H_{s/d}$ is expressed in the momentum space. $H_{T,s/d}$ is the coupling between the source or detection electrode and the infinite ribbon. Here, we assume the electron flow is locally injected at the site $\mathbf{i}_{s}$ in the n region and detected at site $\mathbf{i}_{d}$ in the p region.
For convenience, we define the central scattering region [the solid box in Fig.2(a)] enclosing the injecting site and detecting site. Concerning the central scattering Hamiltonian $H_c$, the total Hamiltonian can also be written in the following form:
\begin{equation}
  H=H_c+\sum_{\beta=s,d,l,r} (H_\beta + H_{T,\beta}) \label{H2}
\end{equation}
Eq.(\ref{H2}) describes a typical open system. Here, we can treat $H_\beta$ as open boundaries, denoting the source electrode, the detection electrode and the left and right semi-infinite lead, respectively.
$H_{T,\beta}$ is the coupling between central scattering region and open boundaries.
Obviously, $H_0=H_c+H_{l}+H_{r}+H_{T}$, and
\begin{equation}
\begin{split}
&H_{l/r}= \sum_{\mathbf{i}{\in l/r}}d_{\mathbf{i}}^{\dag}H_{\mathbf{i}}d_{\mathbf{i}}
+\sum_{<\mathbf{i,j}>{\in l/r}}d_{\mathbf{i}}^{\dag}H_{\mathbf{i,j}}d_{\mathbf{j}}\\
&H_{T}=[H_{l,c}+H_{c,r}]+H.c.\\
&H_{l,c}=\sum_{<\mathbf{i}{\in l},\mathbf{j}{\in c}>}H_x d_{\mathbf{i}}^\dag d_{\mathbf{j}},~
 H_{c,r}=\sum_{<\mathbf{i}{\in c},\mathbf{j}{\in r}>}H_x d_{\mathbf{i}}^\dag d_{\mathbf{j}}
\end{split}
\end{equation}

Next, with the help of the NEGF, the response signals, i.e., the local density $\rho_{\mathbf{i}}$ in the scattering region are calculated as follows. %We first consider the local density $\rho$ at every site $\mathbf{i}$, which is defined as
\begin{equation}
\rho_{\mathbf{i}}=-i\int dE\mathbf{G}_{\mathbf{ii}}^{<}(E)%\nonumber
\end{equation}
where $\mathbf{G}_\mathbf{ii}^{<}$ is the diagonal element of the lesser Green's function.
Using Keldysh equation \cite{Jauho1994}, the lesser Green's function can be written as
\begin{equation}
 \mathbf{G}^{<}=\sum_{\beta=l,r,s,d}\mathbf{G}^{r}i\mathbf{\Gamma}_\beta f_\beta\mathbf{G}^{a}%\nonumber
\end{equation}
Here, $\mathbf{G}^{r}$ and $\mathbf{G}^{a}$ are retarded and advanced Green's function of the scattering region, respectively. $\mathbf{G}^{r}=\mathbf{G}^{a,\dagger}=[E-H_c-\sum_\beta\mathbf{\Sigma}^{r}_\beta]^{-1}$, $f_\beta$ is the Fermi distribution function of the terminal-$\beta$. In the nonequilibrium system, the Fermi energy of the terminal-$\beta$ is shifted by the external bias $V_\beta$, and $f_\beta(E)=f_{0}(E-eV_\beta)$, where $f_0$ is the fermi distribution function with zero bias.
The linewidth function $\mathbf{\Gamma}_{\beta}=i(\mathbf{\Sigma}^r_\beta-\mathbf{\Sigma}^{a}_\beta)$ with $\mathbf{\Sigma}^r_{\beta}$ being the retarded self energy induced by the lead-$\beta$. For the left and right seimi-infinite lead, $\mathbf{\Sigma}^r_{l/r}=\mathbf{H}_{c,l/r}\mathbf{g}^r_{l/r}\mathbf{H}_{l/r,c}$, where $\mathbf{H}_{c,l/r}$ is the coupling from the central region to the left or right lead, $\mathbf{g}^r_{l/r}$ is the surface Green's function of the semi-infinite lead, which can be calculated iteratively using transfer matrix\cite{J.Phys.F:Met.Phys.1984,J.Phys.F:Met.Phys.1985} or Bloch eigenvector.\cite{Lee1981,Lee1981a} The source and detection terminal are expressed in the momentum space. In the wide band limit, the self energy of the source or  the detecting lead is $\mathbf{\Sigma}^r_{s/d}=-i\pi\rho_0 t_{s/d}^2$. Comparing with injecting terminal, the influence of the detection terminal is much weaker, i.e. $t_d\ll t_s$. In this case, we can neglect $\Sigma^{r/a}_d$, then $\mathbf{G}^{r/a}=[E-H_c-\sum_{\alpha}\mathbf{\Sigma}^{r/a}_\alpha]^{-1}$ with $\alpha=l,r,s$.
%Concerning the matrix inverse operation, because of the huge size of the matrix, we have used the transfer matrix technique, through which we can solve the Dyson equation of $G^r$ layer by layer.

In our calculation, the electron flow is injected from source terminal, the left and right semi-infinite lead are all the drain terminals. So, we set $V_s=V$ and $V_l=V_r=0$. Finally, the lesser Green's function can be divided into equilibrium and nonequlibrium term, i.e., $\mathbf{G}^{<}=\mathbf{G}^{<}_{0}+\mathbf{G}^{<}_{V}$ with
\begin{equation}
\begin{split}
&\mathbf{G}^{<}_{0}=\sum_\beta\mathbf{G}^{r}i\mathbf{\Gamma}_\beta\mathbf{G}^{a}f_0\\
&\mathbf{G}^{<}_{V}=\mathbf{G}^{r}i\mathbf{\Gamma}_s\mathbf{G}^{a}(f_s-f_0)%\nonumber
\end{split}
\end{equation}
Here, only the nonequlibrium term contributes to the response signals. It means
\begin{equation}\label{}
\rho_{\mathbf{i}}=\int dE\left[\mathbf{G}^{r}\mathbf{\Gamma}_s\mathbf{G}^{a}\right]_{\mathbf{ii}}(f_s-f_0)\nonumber
\end{equation}
In zero temperature and linear bias limit,
$
\rho_{\mathbf{i}}=\mathbf{G}^{r}_{\mathbf{i},\mathbf{i}_s}\mathbf{\Gamma}_s\mathbf{G}^{a}_{\mathbf{i}_s,\mathbf{i}}eV_s\nonumber
$.
Then, we can define the local partial density
\begin{equation}\label{}
\delta\rho_{\mathbf{i}}/\delta (eV_s) = \mathbf{G}^{r}_{\mathbf{i},\mathbf{i}_s}\mathbf{\Gamma}_s\mathbf{G}^{a}_{\mathbf{i}_s,\mathbf{i}}%\nonumber
\end{equation}

On the other hand, we can also calculate the local conductance, which is defined as $\sigma_{\mathbf{i}}=\partial J_\mathbf{i}/\partial V_s$, where $J_\mathbf{i}$ is the current flowing to the detection terminal that is located at site $\mathbf{i}$. According to the Landauer-B\"{u}ttiker formalism,
\begin{equation}\label{}
J_\mathbf{i}=\frac{e}{h}\mathbf{\Gamma}_{d}
\mathbf{G}^{r}_{\mathbf{i},\mathbf{i}_s}\mathbf{\Gamma}_s\mathbf{G}^{a}_{\mathbf{i}_s,\mathbf{i}}(eV_d-eV_s)\nonumber
\end{equation}
Since we have set $V_d=0$, the local conductance $\sigma_\mathbf{i}$ is then expressed as
\begin{equation}\label{}
\sigma_\mathbf{i}=\frac{e^2}{h}\mathbf{\Gamma}_{d}\mathbf{G}^{r}_{\mathbf{i},\mathbf{i}_s}
\mathbf{\Gamma}_s\mathbf{G}^{a}_{\mathbf{i}_s,\mathbf{i}}%\nonumber
\end{equation}
Here, $\mathbf{\Gamma}_{d}=2\pi\rho_0t_d^2$ is a constant, so the local conductance is equivalent to the partial density, i.e.,  $\sigma_\mathbf{i}\propto \delta\rho_{\mathbf{i}}/\delta (eV_s)$. In the following numerical calculation, only the local partial density is considered.

%Alternatively, the local current density can also be calculated flowing from the site $\mathbf{i}$ to the nearest neighbor site $\mathbf{j}$ can be calculated from the formula \cite{Xing2010,Caroli1971,Jiang2009,Xing2011}
%\begin{eqnarray}
%% \nonumber to remove numbering (before each equation)
%  \mathbf{J_{\mathbf{i}\mathbf{j}}} &=& \frac{e}{h}\int_{-\infty}^{\infty}dE[\mathbf{G}_{\mathbf{i}\mathbf{j}}^{<}\mathbf{H}_{\mathbf{j}\mathbf{i}}-
%  \mathbf{H}_{\mathbf{i}\mathbf{j}}\mathbf{G}_{\mathbf{j}\mathbf{i}}^{<}] \nonumber  \\
%   &=& \frac{2e}{h}\int_{-\infty}^{\infty}dERe[\mathbf{G}_{\mathbf{i}\mathbf{j}}^{<}\mathbf{H}_{\mathbf{j}\mathbf{i}}]
%\end{eqnarray}
%where $\mathbf{G}_{\mathbf{i}\mathbf{j}}^{<}$ is the matrix element of the lesser Green's function of the scattering region that including the source and detector sites.
%
%note that the parts of left and right leads is zero due to $V_{L/R}=0$. Considering a small source voltage and substituting Eq. (9) into Eq. (6), with zero temperature, the current-density vector can be written as
%\begin{equation}\label{}
%  \mathbf{J_{ij}}=\frac{2e^2}{h}Re\{[i\mathbf{G}^r\mathbf{\Gamma}_s\mathbf{G}^a]_{\mathbf{i}\mathbf{j}}\mathbf{H}_{\mathbf{j}\mathbf{i}}\}\times\delta V
%\end{equation}

\section{NUMERICAL RESULTS AND DISCUSSION}

In the numerical calculation, the parameters of 3DTI are set as $D_{0}=0.28eV$,\cite{Liu2010} $D_{\perp}=10eV{\AA}^2$, $D_{\parallel}=56.6eV{\AA}^2$, $\epsilon_{k}=0$, $A_{\perp}=2.2eV{\AA}$, $A_{\parallel}=4.1eV{\AA}$,\cite{Zhang2009} the lattice constant $a=5{\AA}$. Here we set $E_{F}=0$, so the (kinetic) energy (relative to the energy of $\Gamma$ point) of electrons and holes are $E_n=E_0$ and $E_p=-E_0$, respectively.

\subsection{Focusing effect in the linear regime}

\begin{figure}
\includegraphics[width=8.7cm,clip=]{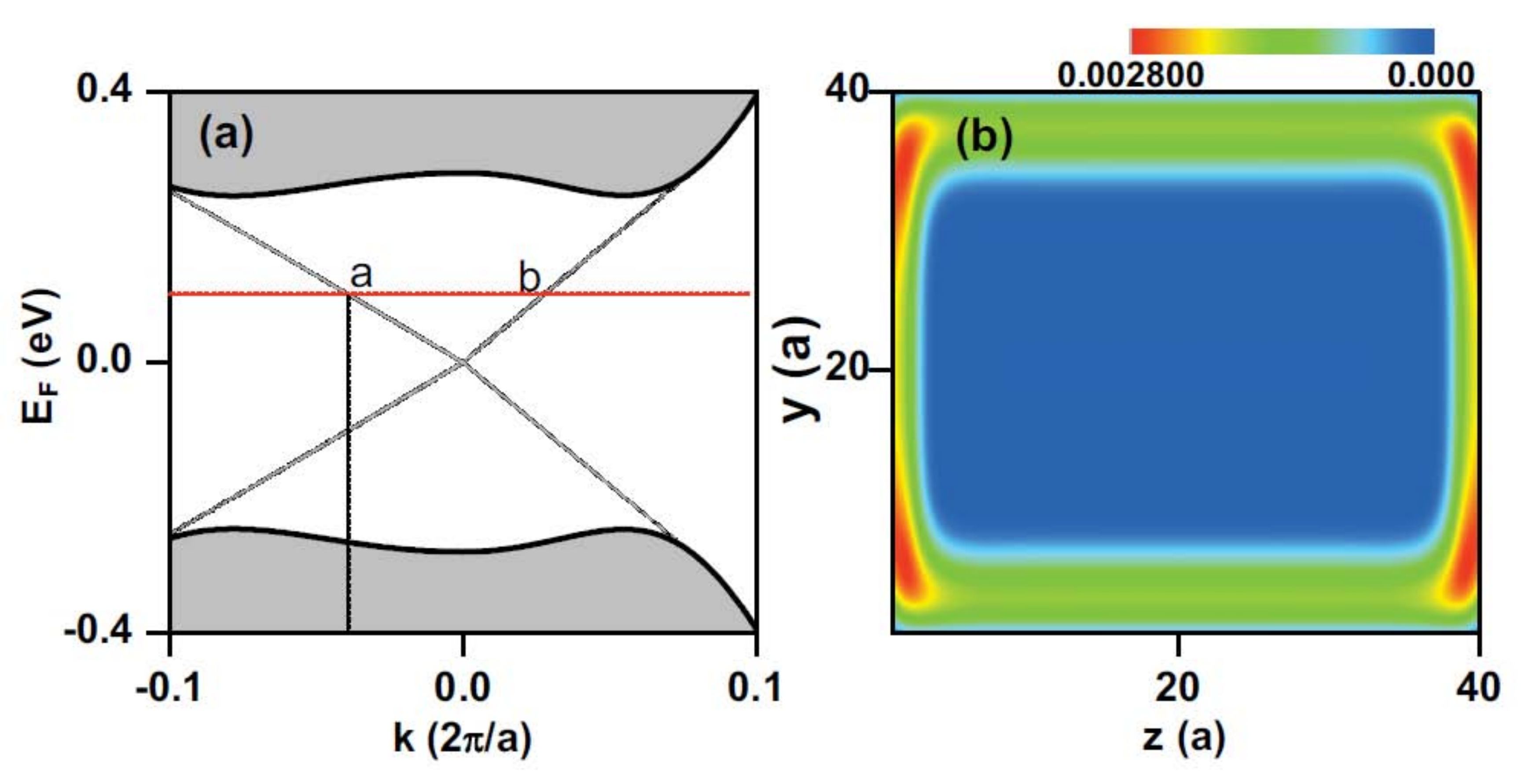}
\caption{ (Color online) Panel (a): local density of state at the low energy regime on an infinite X-Y
surface of a semi-infinite 3DTI along the direction of $\Gamma\rightarrow K$ and $\Gamma\rightarrow M$. The black thick lines are the bulk band edge. The red dotted lines denotes the fixed low energy $E_F=0.1eV$ and corresponding $k_x=0.04$, with which the local density of state in the cross finite ($40\times40$) Y-Z section of infinite ribbon are plotted in panel (b).}
\end{figure}

For the 3D TI, because of the band inversion near $\Gamma$ point, the nontrivial energy gap is induced.
When $A_\parallel$ is small, the energy gap is roughly determined by $D_0$.
In our model, the global bulk energy gap is about in the interval of [-0.245eV,0.245eV].
When the incident energy of electron is within the bulk energy gap, all the bulk states are forbidden,
only the linear massless Dirac Fermions on the surfaces are permitted. In Fig.3, we first study the equilibrium density of state $\rho_0$ in this energy regime. According to fluctuation-dissipation theorem, $\rho_0= \frac{i}{2\pi}[G^r_s-G^{r,\dagger}_s]$, where $G^r_s$ is the surface Green's function of semi-infinite 3D TI for fixed Fermi energy. In Fig.3(a) we plot the distribution of $\rho_0(k,E_F)$ of infinite surface along $x$-$y$ plan.
The infinite surface is denoted by 2D momentum $k_{x,y}$. In order to show the result intuitively, the momentum is set along the line of $\overrightarrow{\Gamma M}$ ($k_x=k_y=k$) and $\overrightarrow{\Gamma K}$ ($k_x=k,k_y=0$), where $\Gamma=(0,0)$, $M=(-0.5,0)$ and $K=(0.5,0.5)$ are all the high symmetry points in momentum space. From Fig.3(a), we can clearly see the linear dispersion of the massless Dirac cone within the bulk energy gap that is bordered by the thick black lines in Fig.3(a). Fixing $E_F=0.1eV$, the corresponding momentum $k_x$ along the path of $\overrightarrow{\Gamma M}$ is determined [the black dotted line in Fig.3(a)]. With this $E_F$ and $k_x$, the local density of state in the finite cross section ($40a\times40a$) of infinite ribbon is plotted in Fig.3(b).
From Fig.3(b), we can clearly see the boundary states (the red region along the boundary) in the $y$-$z$ section, depicting the surface states in the infinite ribbon.

\begin{figure}
\includegraphics[bb=50mm 53mm 250mm 154mm,width=8.7cm,totalheight=4.4cm, clip=]{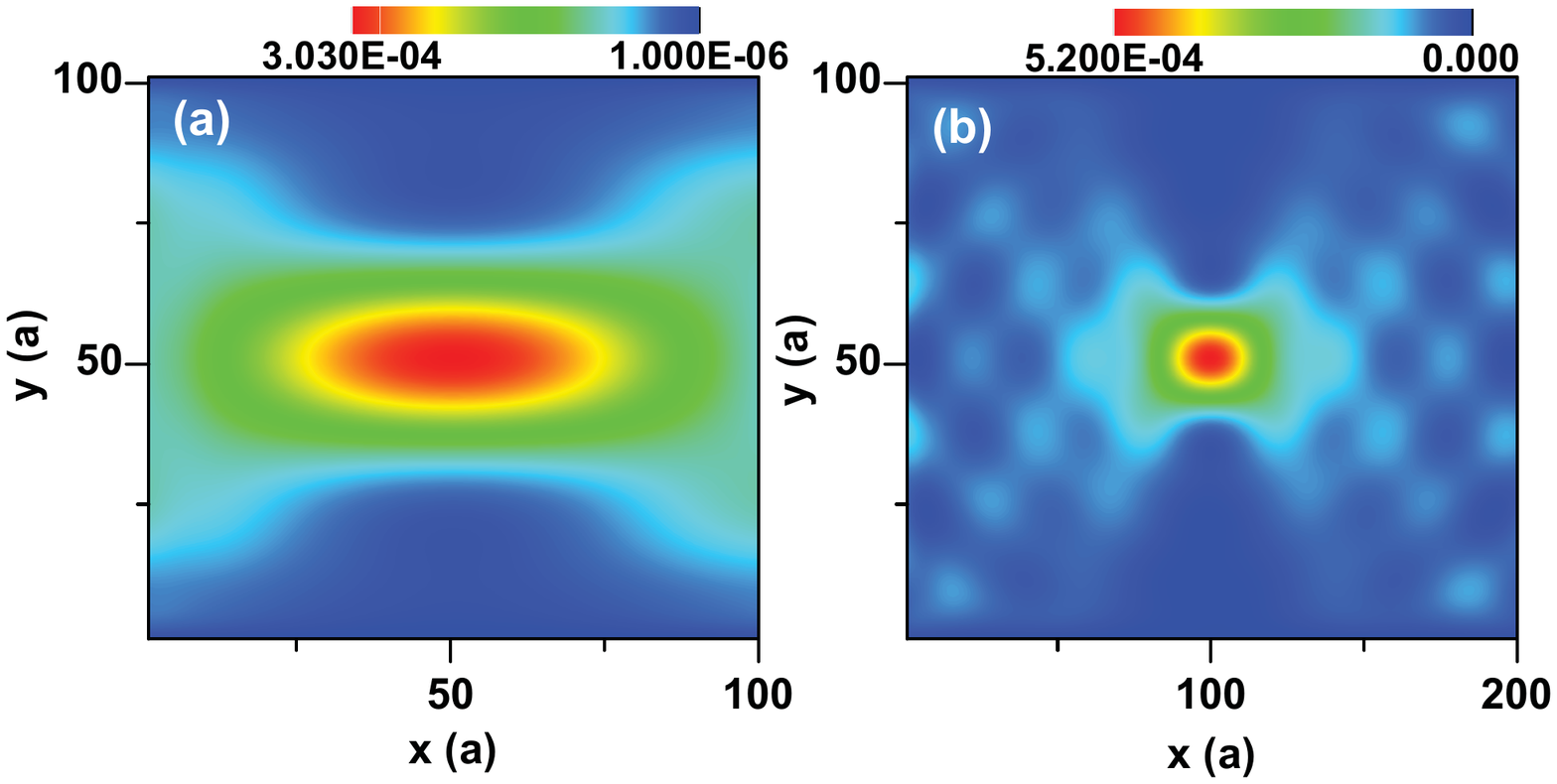}
\caption{ (Color online) The focusing effect dominated by surface states for $E_{0}=0.1eV$ and $0.2eV$.
Panel (a): distribution of local partial density at the bottom layers ($z=0$) in the p region while electron flow is injected from the position $(-49.5a,50a,0)$ in the n region. The potential height of PNJ $E_{0}=0.1eV$. Panel (b): $E_0=0.2eV$, the electron flow is injected from $(-99.5a,50a,0)$ and focused at $(99.5a,50a,0)$. The width of nanoribbon $W_y=100a$, the height of nanoribbon $W_z=6a$.}
\end{figure}

From Fig.3, we have confirmed the surface states in both momentum space and real space when the incident energy is within the bulk energy gap. In the following, we will study the focusing effect induced by these surface states.
In Fig.4, we focus on the focusing effect in surfaces of 3D TI nanoribbon with straight PNJ potential within bulk energy gap. In Fig.4(a) and Fig.4(b), we plot the distribution of local partial density $\delta \rho/\delta eV$ at the bottom layer in the p region ($x>0$) for $E_0=0.1eV$ and $0.2eV$, respectively. The results in other surfaces are similar (not shown). The width and height of ribbon are set as $W_y=100a$, $W_z=6a$, respectively.
The local partial density on bottom surface is the sum of the two layers at the lowest bottom, i.e., the layers at $z=0.5a$ and $z=1.5a$. In the panel (a), electron flow is injected from the position $(-49.5a,49.5a,0.5a)$ in the n region, and focused at the position $(49.5a,49.5a,0.5a)$ in the p region.
While in the panel (b), the electron flow is injected from $(-99.5a,49.5a,0.5a)$ and focused at $(99.5a,49.5a,0.5a)$.
It can be seen as long as the energy is within the energy gap, the electron flow can be perfectly focused at the symmetric position in the p region, no matter where it is injected in the n region, as analysed in the Fig.1.
It is similar as in graphene\cite{Cheianov2007a,Xing2010}. Besides, we also find that the focusing effect for the higher energy ($E_0=0.2eV$) is better, which is totally different from the focusing effect in graphene, for which the focusing effect is worse for the higher energy\cite{Xing2010}. In addition, due to the extra scattering induced by the boundaries of the nanoribbon, there are regular interference patterns when the scattering region is long, as shown in the Fig.4(b).

\begin{figure}
\includegraphics[width=7.0cm, clip=]{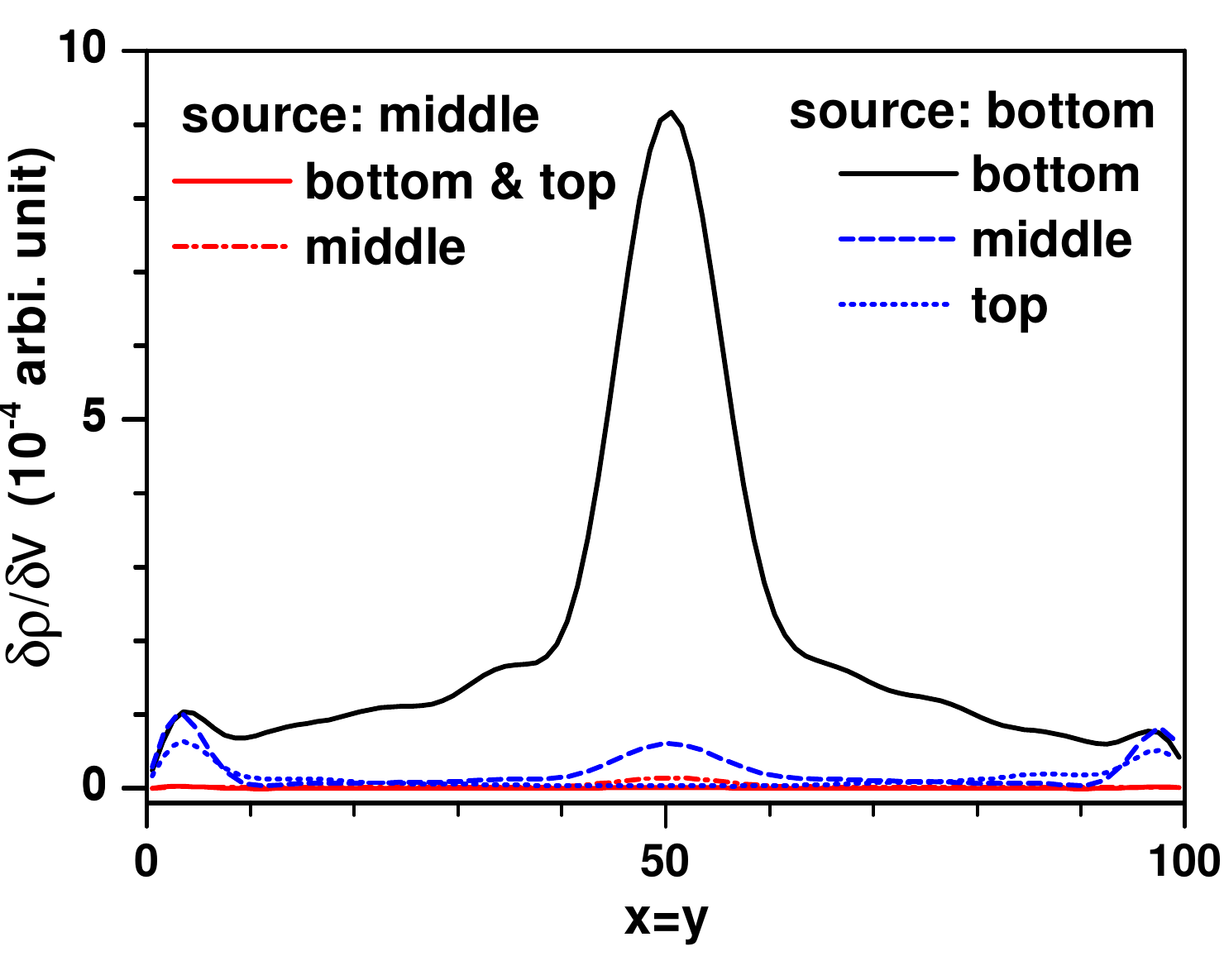}
\caption{ (Color online) Distribution of local particle density along the diagonal line of the p region, i.e., $(x=y)\in[0,100a]$ for injecting energy $E_{0}=0.2eV$ (surface states dominated focusing).
Electron flow is injected from the middle layer ($z=6.5a$) or bottom layer ($z=0.5a$).
The width of ribbon $W_y=100a$, the height of ribbon $W_z=13a$.}
\end{figure}

In the linear regime, the focusing effect is dominated by the surface states, so the focusing effect can't happen in the deep of bulk. In Fig.5, considering a nanoribbon with size $W_y=100a$ and $W_z=13a$, the distribution of local particle density $\delta\rho/\delta eV$ along the diagonal line of the p region, i.e., $(x=y)\in[0,100a]$, is plotted. Assuming the electron flow is injected from the middle layer ($z=6.5a$, red lines) and bottom layer ($z=0.5a$, black and blue lines), we plot $\delta\rho/\delta eV$ at bottom, middle and top layers in the p region.
Here, $\delta\rho/\delta eV$ of the middle layers is the sum of the five middle layers, i.e., the layers located at $z$=4.5-8.5$a$. $\delta\rho/\delta eV$ of the bottom (top) layers is the sum of the four lowest (highest) layers, i.e., the layers located at $z$=0.5-3.5$a$ (9.5-12.5$a$). It is found when injecting electron flow from middle layer, the local response in the p region is uniformly small in all layers, which means no focusing effect happens when electron source is located deep inside the bulk. On the other hand, when injecting electron flow from the bottom layer, $\delta\rho/\delta eV$ increases abruptly in the center of the bottom layers (see the black line in Fig.5), and becomes very small in the middle and top layers (the blue lines). In a word, in the low energy linear regime, the focusing effect of electron flow is dominated by the surface states and arises only in the surfaces of 3D TI ribbon.

\subsection{Focusing effect in high energy regime}

\begin{figure}
\includegraphics[width=8.5cm,clip=]{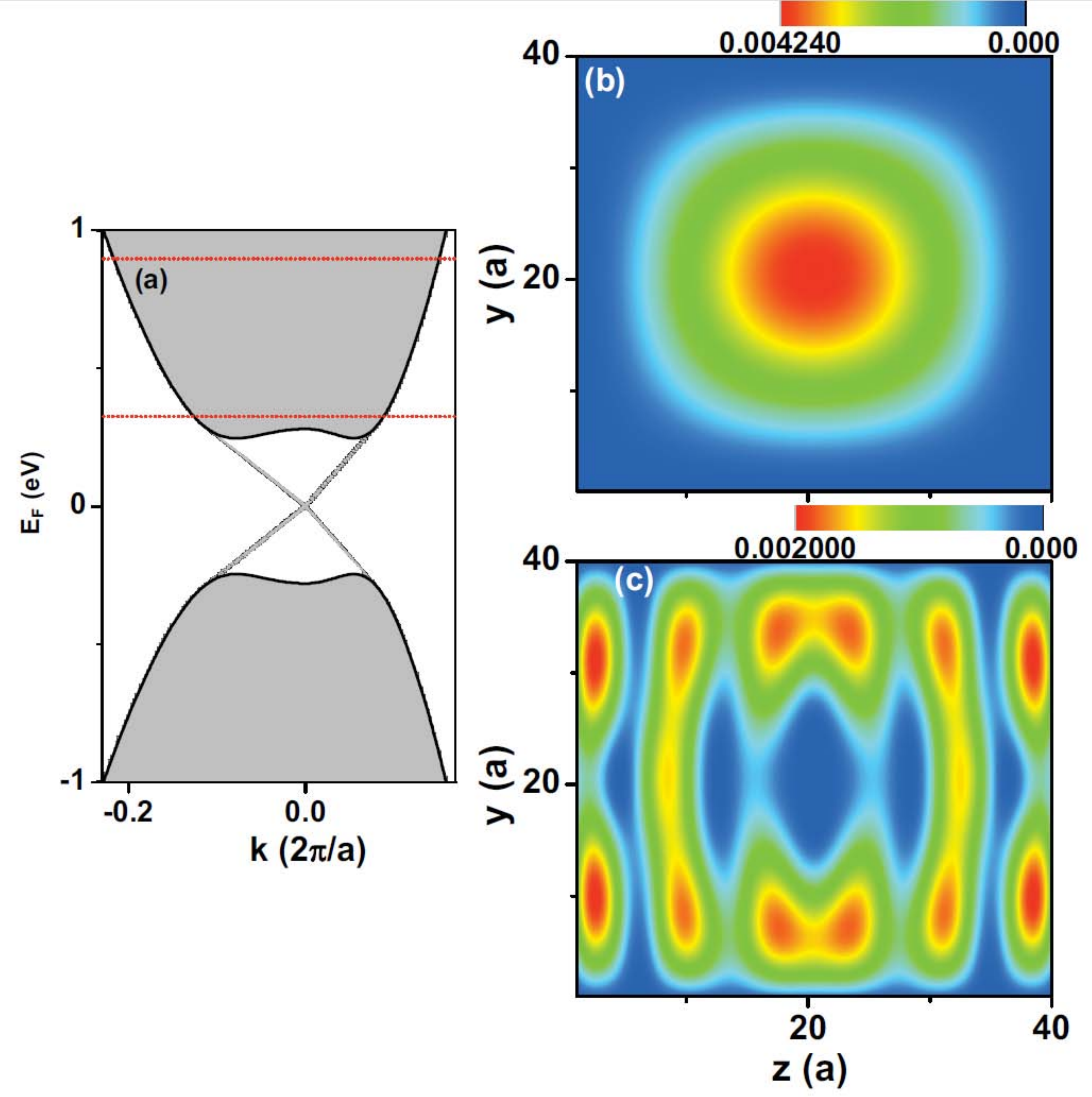}
\caption{ (Color online) local density of state at high energy regime for an infinite X-Y
surface of a semi-infinite 3D TI. The momentum $k$ is along the direction of $\Gamma$-$M$ and $\Gamma$-$K$.
The red dotted lines denotes the fixed Fermi energy $E_F=0.33eV$ and $0.9eV$, which corresponds to the puddle regime of surface and bulk states, and the bulk states dominated regime, respectively. Corresponding to $E_F=0.9eV$ and $0.33eV$, the local density of state in the cross finite ($40\times40$) Y-Z section of infinite ribbon are plotted in panel (b) and (c) respectively.}
\end{figure}

When the incoming Fermi energy is beyond the bulk energy gap, carriers are no longer described by linear Dirac cone.
How about the focusing effect in this case? In Fig.6(a), we show the distribution of equilibrium density of state $\rho_0(k,E_F)$ in whole energy regime for an infinite surface of a semi-infinite 3D TI. %along the direction of $\overrightarrow{\Gamma M}$ and $\overrightarrow{\Gamma K}$.
Both the discrete surface states (gray lines) and continuous bulk states (gray region bordered by black band edges) appear in Fig.6(a). Two cases are considered: Fermi energy is near band edge ($E_F=0.33eV$) and deep in conductance band ($E_F=0.9eV$) [see the red dotted lines in Fig.6(a)]. With these $E_F$, the local density of state in the finite cross section ($40a\times40a$) of infinite ribbon is plotted in Fig.6(b) and Fig.6(c). We can see near the band edge [$E_F=0.33$eV, panel (b)] the surface states of the infinite ribbon are disturbed meanwhile the bulk states have not yet predominated. When $E_F$ is deep in conductance band [$E_F=0.9$eV, panel (c)], the surface states disappear completely. Then, the system is dominated by the bulk states that do not obey the massless linear Dirac equation. In the following we will study the focusing effect induced by the bulk states.

\begin{figure}
\includegraphics[width=8.05cm, clip=]{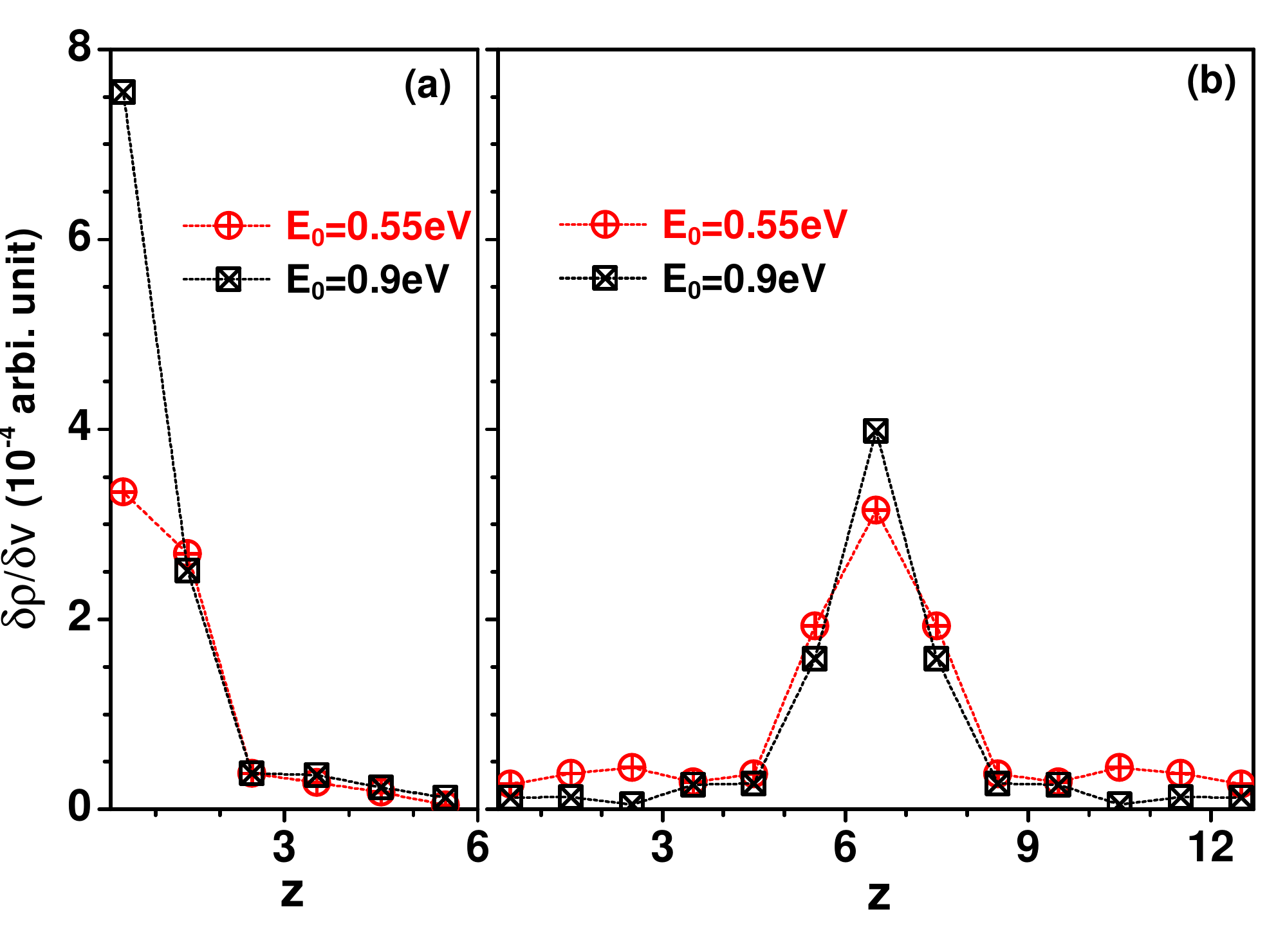}
\caption{ (Color online) In the high energy regime, $E_0=0.55eV$ and $0.9eV$, the $\delta\rho/\delta V$ at the center ($x=0.5W_x$, $y=0.5W_y$) of the p region vs $z$. Panel (a): $W_z=6a$, electron flow is injected from the bottom layers ($z=0.5a$). Panel (b): $W_z=6a$, electron flow is injected from the middle layer ($z=6.5a$).}
\end{figure}

We first set PNJ potential $E_0\gg D_0$. In this case, the bulk states dominate the transport processes.
As a result, the focusing effect can occur in bulk as well as the surfaces, which is different from the focusing effect in the linear energy regime. Injecting electron flow from the bottom layer located at $z=0.5a$, we plot the local partial density $\delta\rho/\delta eV$ in the center of p region at every layer signed by its $z$ coordinate in Fig.7(a). The height of the nanoribbion is $W_z=6a$. It can be seen $\delta\rho/\delta V$ becomes maximum at bottom $x$-$y$ plane with $z=0.5a$ since electron source is located in bottom layer. When deviating from the bottom layer, $\delta\rho/\delta eV$ reduces abruptly to nearly zero. Next, we set $W_z=13a$ and inject electron flow from the middle layer ($z=6.5a$). $\delta\rho/\delta eV$ in the center of p region at every layer is plotted in the Fig.7(b). We can see  $\delta\rho/\delta eV$ becomes maximum at the middle bulk layer ($z=6.5a$) and decreases abruptly at other layers. In other words, wherever injected from the n region, the electron flow can always be perfectly focused to the site $(x,y,z)$ in the p region. It means except the 2D linear massless Dirac Fermions, the conventional semi-metal can also produce the focusing effect.

\begin{figure}
\includegraphics[bb=50mm 53mm 250mm 154mm,width=8.8cm, clip=]{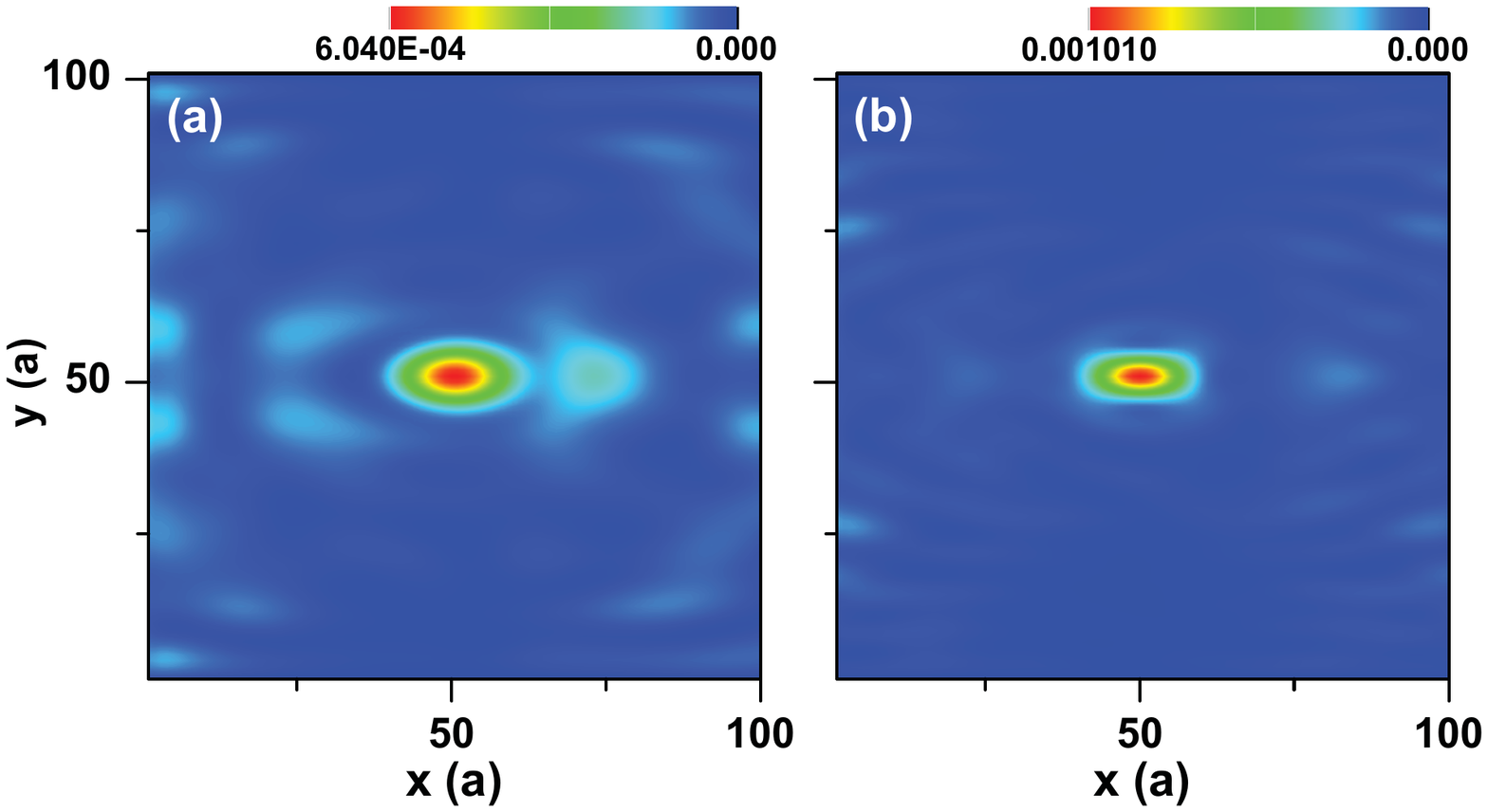}
\caption{ (Color online) The focusing effect dominated by bulk states.
Panel (a): distribution of local partial density in the p region of the infinite ribbon with a sharp PNJ for $E_{0}=0.55eV$ panel (b): $E_{0}=0.9eV$. The other parameters: the width $W_y=100a$, the height $W_z=6a$.}
\end{figure}

\begin{figure}
\includegraphics[bb=8mm 50mm 230mm 250mm,totalheight=6cm, clip=]{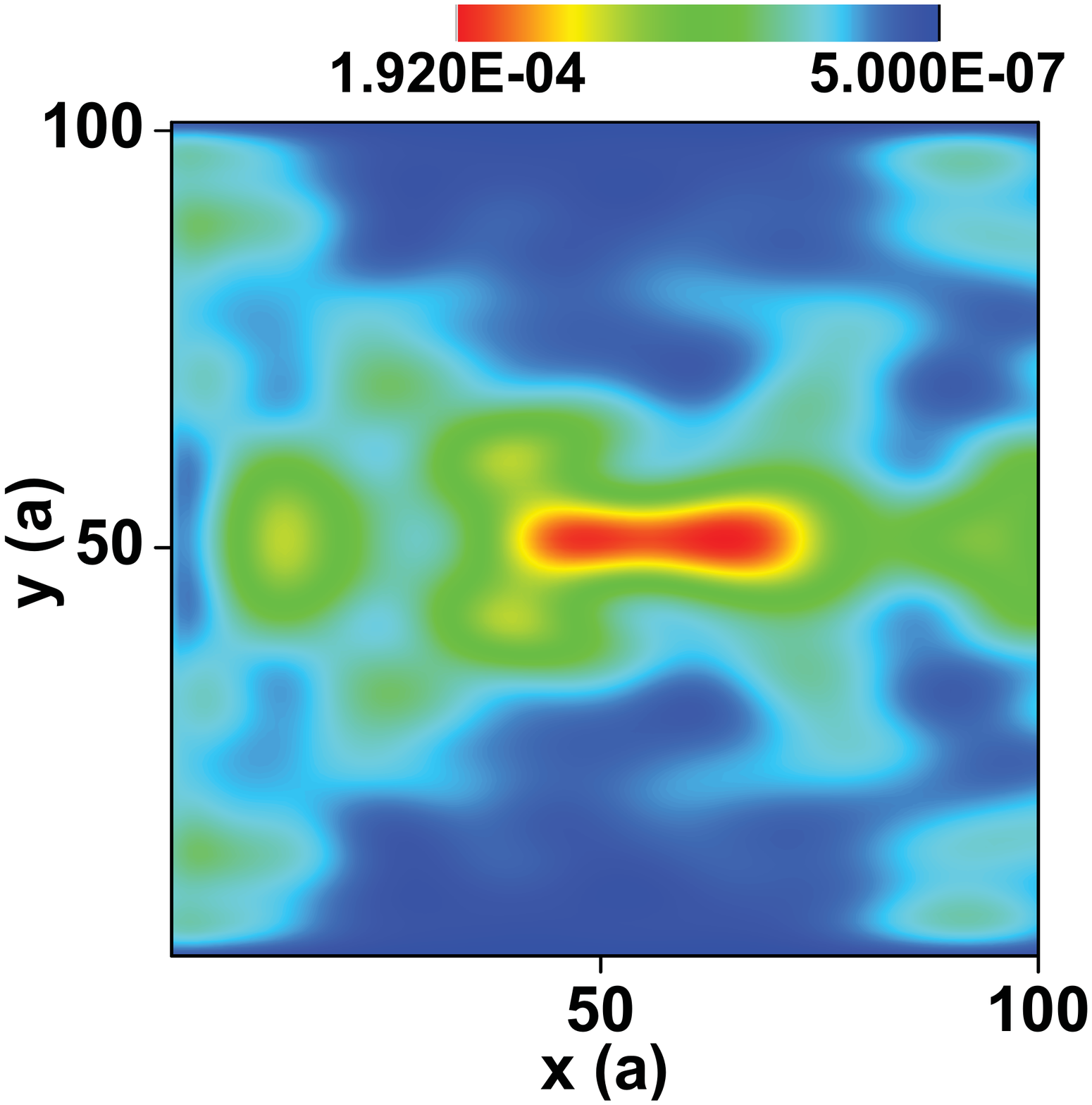}
\caption{ (Color online)  Distribution of local partial density of the p region in the puddle regime of surface and bulk states for $E_{0}=0.33eV$. The other parameters are same as Fig.8.}
\end{figure}

In general, it is difficult to embed the source lead deep into the bulk. So, in the following, the source terminal is assumed to be located in the bottom surface. Of course, the focusing effect occurs on the bottom surface as well. However, we must keep in mind that this focusing is dominated by bulk states, not by surface states.
In Fig.8, we plot the distribution of local partial density $\delta\rho/\delta eV$ in the p region for $E_0=0.55eV$ and $0.9eV$ that are all deep inside the bulk energy band. Here, the local partial density in the bottom surface is the sum of the two lowest layers ($z=0.5a$ and $1.5a$). From Fig.8, we can see the perfect focusing effect in the high energy regime. The higher the PNJ potential, the better the focusing effect is. Different from the Fig.4, this focusing effect is induced by the conventional bulk states with quasi-quadratic dispersion.

Up to now, we have shown that both the surface states (with linear dispersion) and the bulk states (with quadratic dispersion) can produce focusing effect. %Next, we will study the focusing effect near the band edge.
Now, we wonder if the focusing effect is going to happen  when the surface states and the bulk states are mixed.
In Fig.9, we plot the distribution of local partial density $\delta\rho/\delta eV$ in p region for $E_0=0.33eV$.
When $E_0=0.33eV$, the Fermi energy is near band edge, surface sates and bulk states coexist as shown in Fig.6(b).
From Fig.9, we can see the focusing effect induced by the mixed states is really much worse compared to the Fig.4 or Fig.8 in which the pure surface states or bulk states are dominant. It is not strange because the dispersion of surface states and the bulk states are different, the mixture states can't synchronously penetrate PNJ through the Klein tunneling\cite{Cheianov2007a}. The poor focusing in Fig.9 just reveals the different behaviors between the surface states and bulk states.

\subsection{Influence of disorder on focusing effect}

In the real device, disorder is inevitable. In this subsection, we will study the influence of disorder.
In general, the impurities may appear near the interface due to the preparation of the PNJ.
Disorders induce random scattering which is simulated by the random on-site potential\cite{Xing2011a,Chen2012a} $\delta\epsilon_{\mathbf{i}}$ uniformly distributed in the interval [$-w/2,w/2$], where $w$ is the disorder strength.
Due to the disorder, the on-site energy now becomes $\epsilon_\mathbf{i}+\delta\epsilon_\mathbf{i}$.
In the numerical calculation, disorders are distributed near the PNJ in the region from $x=-5.5$ to $x=5.5a$.
The numerical results are averaged over 200 random configurations.

\begin{figure}
\includegraphics[bb=8mm 50mm 208mm 245mm,width=8.7cm,clip=]{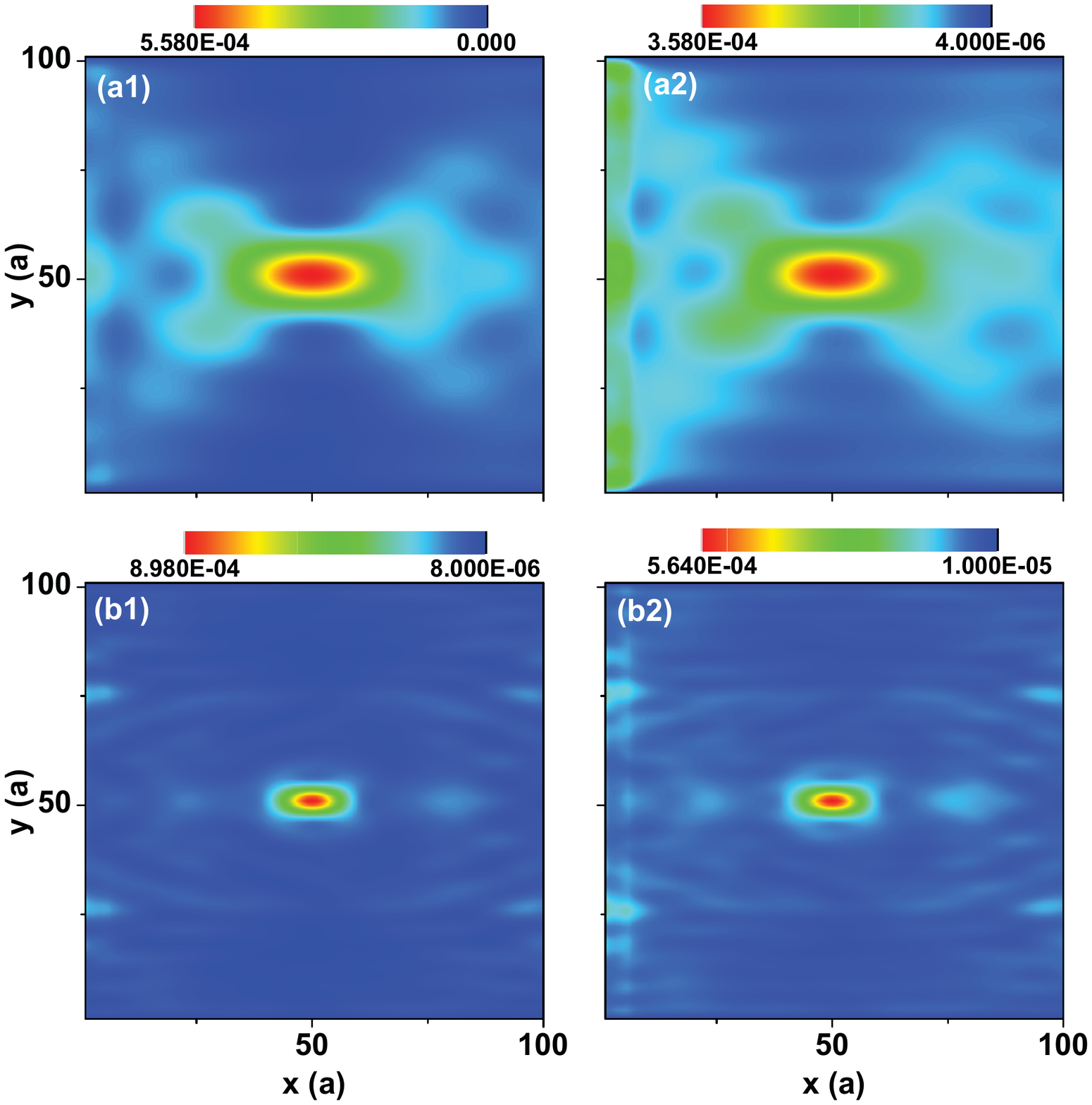}
\caption{ (Color online) Distribution of local partial density of p region in the presence of disordered p/n junction.
The top and bottom panels are corresponding to the surface states dominated regime, $E_{0}=0.2$ for $w=2eV$ and $4eV$ [panel (a1) and (a2)] and bulk states dominated regime, $E_{0}=0.9$ for $w=2eV$ and $3eV$ [panel (b1) and (b2)], respectively. The other parameters are same as Fig.8.}
\end{figure}

\begin{figure}
\includegraphics[width=7.5cm, clip=]{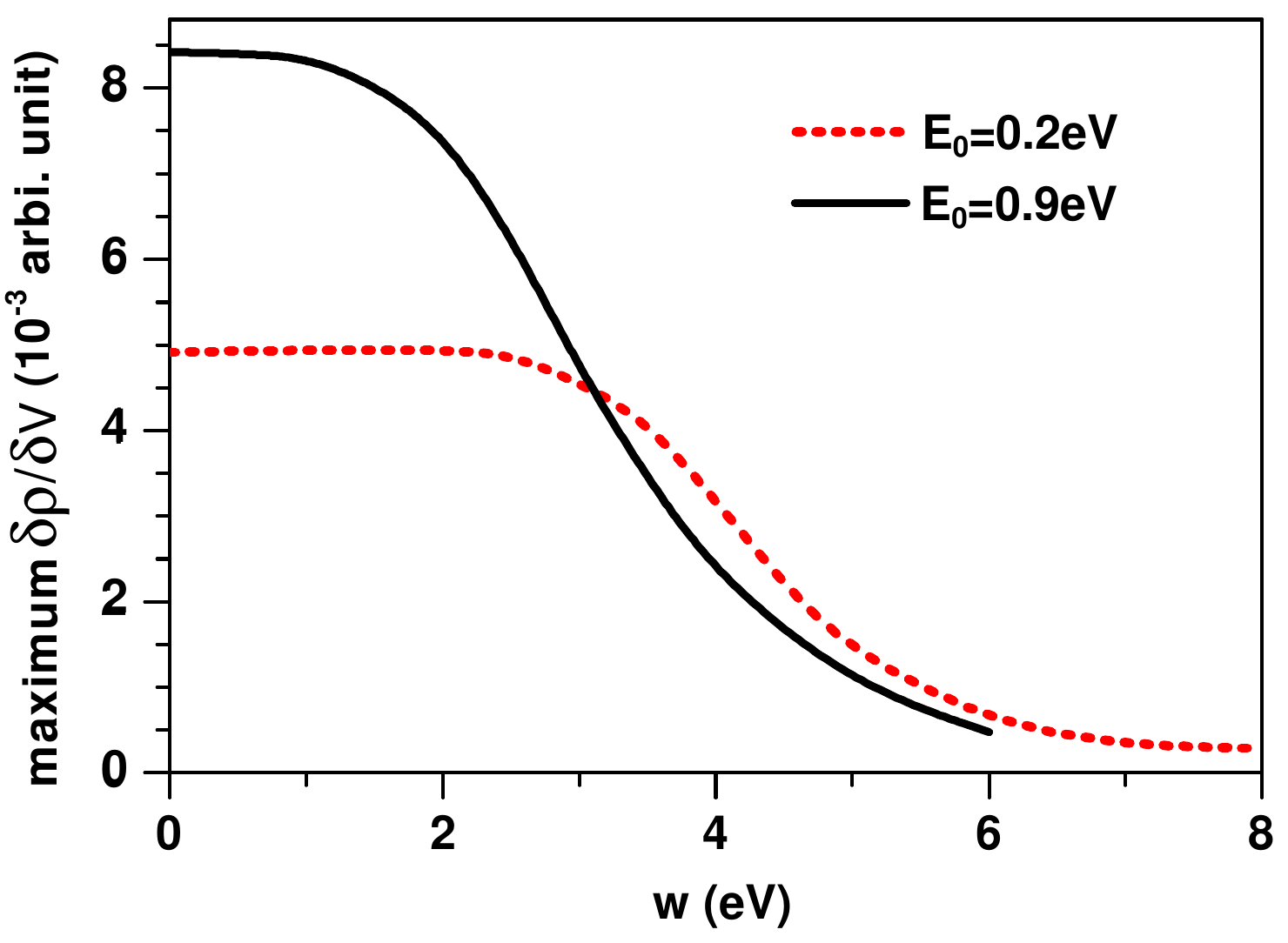}
\caption{ (Color online) The maximum value of local partial density in the p region vs random disorder strength $w$ in surface sates dominated regime, $E_{0}=0.2$ and bulk states dominated regime, $E_{0}=0.9$.}
\end{figure}

In Fig.10 and Fig.11, we study the focusing effect in the presence of random disorder. Both the surface states dominated focusing effect ($E_0=0.2eV$) and bulk states dominated focusing effect ($E_0=0.9eV$) are studied.
In Fig.10(a) and Fig.10(b), we plot the distribution of local partial density for $E_0=0.2eV$ and $E_0=0.9eV$, respectively. The disorder strength is set as $w\geq2eV$ that is very strong comparing to the PNJ potential $E_0$.
From Fig.10, we can see in the presence of strong disorders focusing effects dominated by both the surface states and the bulk states survive successfully. Comparing Fig.10(b) with Fig.8(b), it can be found in the presence of strong disorders the patterns of $\delta\rho/\delta eV$ distribution are hardly affected. So, the focusing effect is quite immune to random disorders, no matter it is dominated by surface states or bulk states. It means the robust focusing effect is a general phenomena, not limited to the massless Dirac Fermions. It is promising for the future device designs.

Comparing Fig.8(b) and Fig.10(b), we can see that although the focusing patterns are kept well, the focusing intensity, i.e., the maximum $\delta\rho/\delta V$, is reduced severely in the presence of strong disorder.
In Fig.11, we plot the maximum value of local partial density vs random disorder strength $w$. Here the maximum value is the sum of local partial density of the nine sites around the central focus (located at$x=49.5a,y=49.5a$).
From Fig.11, we can find for the bulk states dominated focusing ($E_0=0.9eV$), the intensity is maintained when $w<E_0$. When the disorder is strong enough to destroy the PNJ, the maximum $\delta\rho/\delta V$ decreases rapidly, until the focusing patterns are finally smeared. Correspondingly, the focusing effect induced by the PNJ is out of work. For the surface states dominated focusing ($E_0=0.2eV$), the maximum $\delta\rho/\delta V$ remain unchanged even in very strong disorder ($w\gg E_0$) because of the topological nature of the surface states. The focusing effect is kept until the disorder is strong enough to destroy the topological surface states. In summary, surface states dominated focusing effect is coarse but more robust than bulk states dominated focusing effect. In other words, the bulk states dominated focusing effect is finer but frangible comparing to surface states dominated focusing effect.

\subsection{Effect of magnetic field on focusing}

\begin{figure}
\includegraphics[bb=50mm 53mm 250mm 154mm,width=8.8cm, clip=]{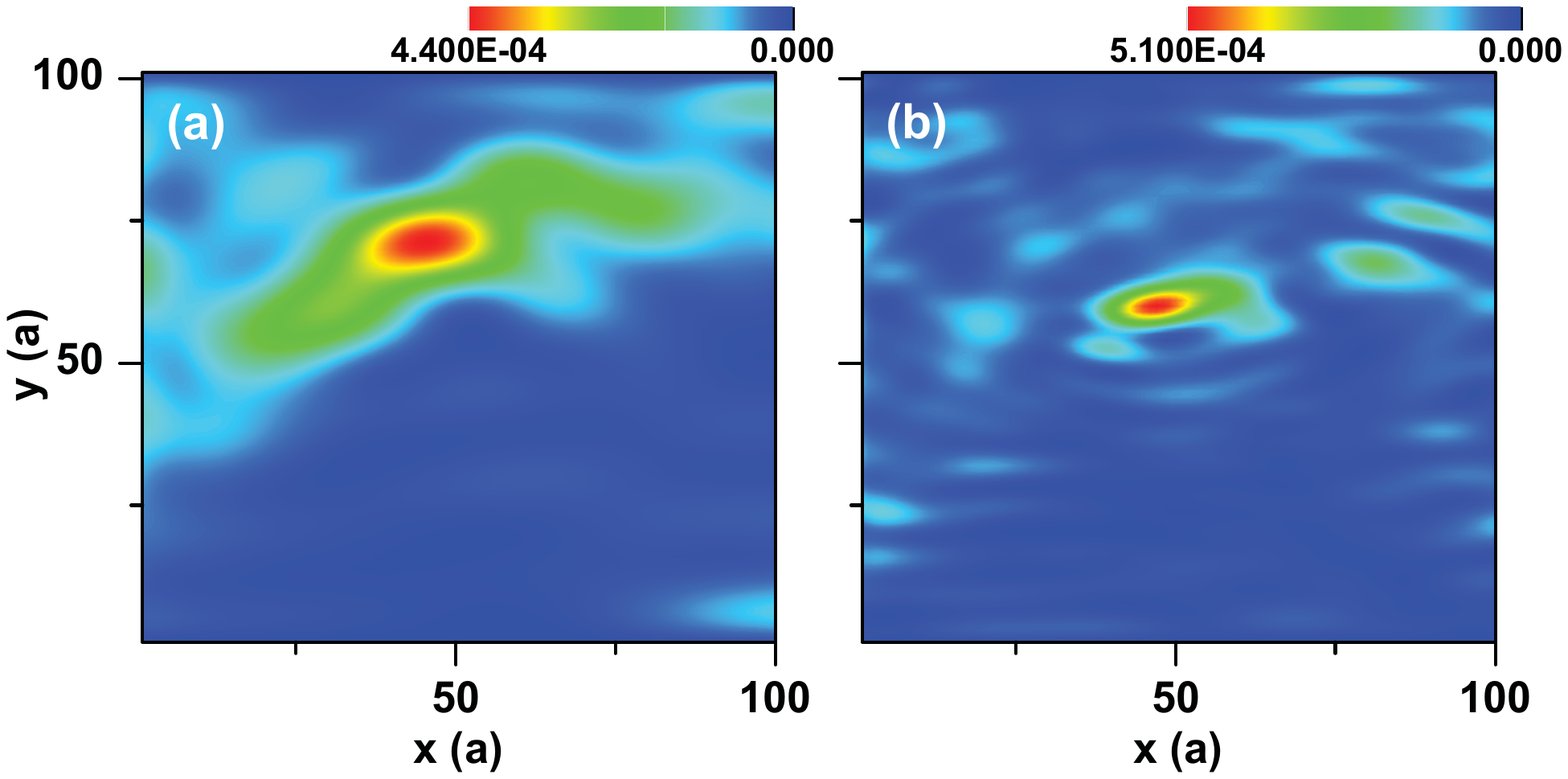}
\caption{ (Color online) Distribution of local partial density of p region in the presence of external magnetic field.
panel (a): surface states dominated regime, $E_{0}=0.2$ panel (b): bulk states dominated regime, $E_{0}=0.9$.
The magnetic filed $B_z=4T$.}
\end{figure}

\begin{figure}
\includegraphics[width=7.5cm, clip=]{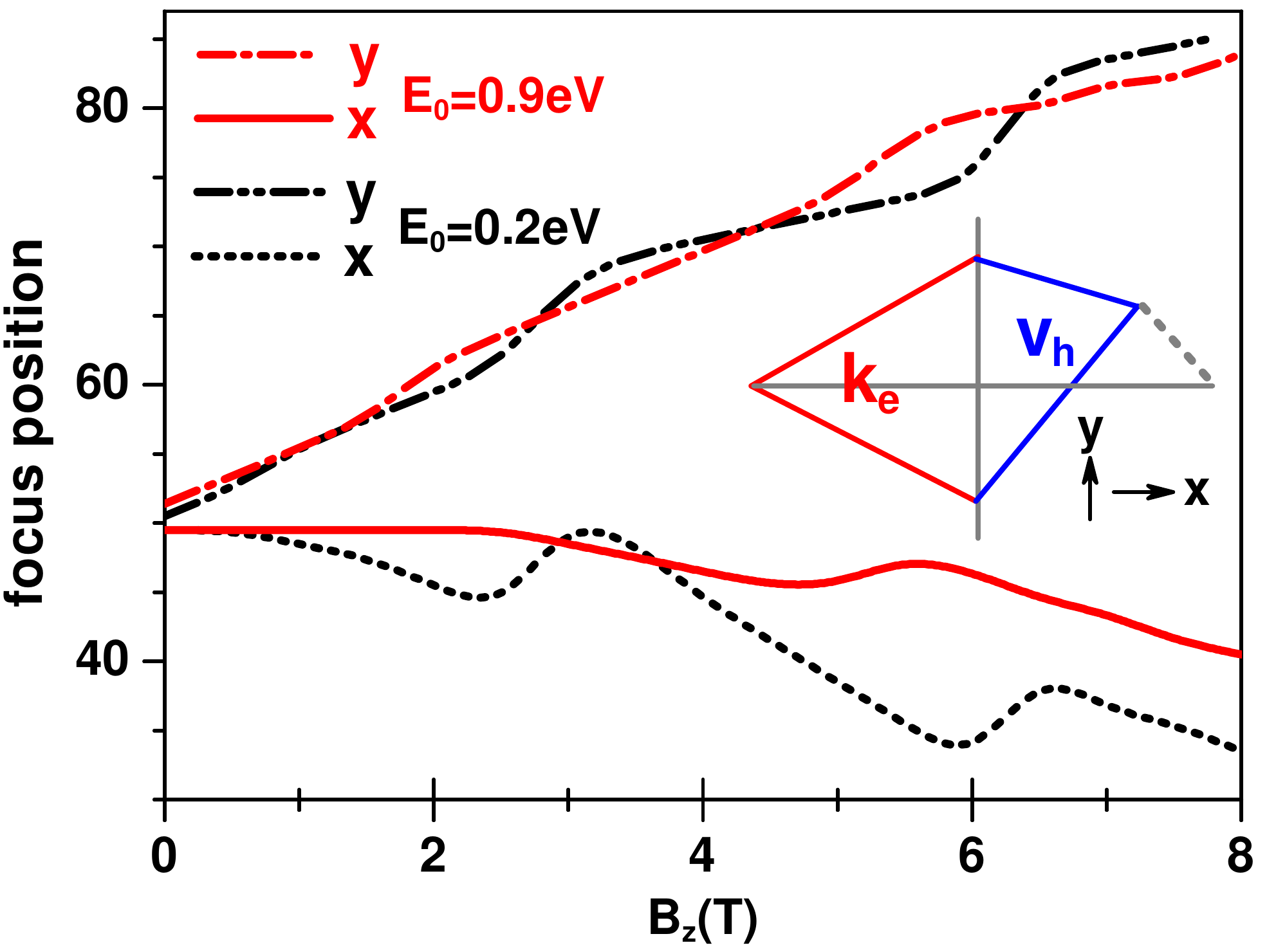}
\caption{ (Color online) Position coordinate of the focused electron flow vs magnetic field $B$ for $E_{0}=0.2eV$ [the black lines] and $0.9eV$ [the red lines], respectively.}
\end{figure}

Besides the random disorders, the focusing effect is disturbed by external magnetic field. In the following, we will study the influence of the perpendicular magnetic field $B_z$. In Fig.12(a) and Fig.12(b), we plot the distribution of local partial density for $E_0=0.2eV$ and $0.9eV$, respectively. The magnetic field is set as $B_z=4T$. It can be seen the focusing pattern is disturbed and the focus point deviates from the central position of p region. Furthermore, comparing Fig.12(b) and Fig.8(b), we can see the focusing intensity is also reduced by $B_z$. This is because the magnetic vector potential produces additional transverse velocity that breaks the conservation of momentum $k_{y}$.
As a result, the symmetric focusing process as analysed in the Fig.1 falls down, the focusing intensity is then weakened. Although weakened by the magnetic field $B_z$, the focusing effect still exist in the p region.
From Fig.12, we can clearly seen the deflected focus point.

Let's analyse how does the magnetic field affect the focusing effect.
In the presence of magnetic field $B_z$, the concomitant vector potential induces anomalous transverse velocity $\delta v_y$.\cite{Rev.Mod.Phys.82.1959-2007.Xiao.2010} In our model, $\delta v_y$ is positive (negative) for electrons (holes). Assuming the electrons are injected in the middle of the ribbon, the momentum $k_e$ is symmetrically distributed, the velocity of holes $v_h$ is shifted by positive $\delta v_y$ as shown the inset of Fig.13.
Then we can expect that the focus point (the blue point) is deflected. It deviates from the center of p region with negative $\delta x$ and positive $\delta y$ (gray dotted line in the inset of Fig.13).
To track the deflection of focus point in the presence of magnetic field, we plot $x$ and $y$ coordinate of the focus position vs $B_z$ in the main panel of Fig.13. It can be seen with the increasing $B_z$ focus point deviates from the central point ($x=49.5a,y=49.5a$) of p region. It is true no matter the focusing effect is dominated by surface states ($E_0=0.2eV$) or bulk states ($E_0=0.9eV$). The deflections of focus point are roughly the same for $E=0.2eV$ and $E=0.9eV$. Besides, considering that the the central scattering region is finite, due to the finite size effect the track of the deflection of focus point oscillates periodically with the oscillating period of $\Delta B\approx\frac{h/e}{W_xW_y}$. This is because when the magnetic flux is quantized by $h/e$ ($h$ is Planck constant and $e$ is electron charge), anomalous transverse velocity $\delta v_y$ becomes maximum and $\delta x$ or $\delta y$ is correspondingly maximum. Furthermore, the oscillation is more violent for lower $E_0$ because the subbands are more discrete.

 \section{CONCLUSION}

In summary, based on tight binding effective Hamiltonian and non-equilibrium Green's function technique, the focusing effect of electron flow in 3D TI with single PNJ is studied.
It is found the electrons/holes flow injecting from n/p region can be perfectly focused at the symmetric position in the p/n region, regardless whether the incident energy is within or beyond the bulk energy gap.
It means in a 3D TI, the focusing effect can be produced by both surface states and bulk states effectively.
So, The focusing effect in a PNJ is a general phenomena, not limited in the massless linear Dirac cones.
Although the focusing effect is absolutely supported by both surface and bulk states, it can not appear when above two type of states are mixed (near the band edges), because of the incompatible dispersion for surface states and bulk states. Furthermore, we also study the influence of random scattering and the weak external magnetic field $B_z$ on the focusing effect. It is found the focusing effect is immune to random disorders. Finally, in the presence of a weak perpendicular magnetic field $B_z$, the focusing effect remains, except the focus point is deflected by the transverse Lorentz force. The discoveries are beneficial for us to get a better insight on the topological materials.

$${\bf ACKNOWLEDGMENTS}$$
This work is supported by the grant of Trans-Century Training Programme Foundation for the Talents by the State Education Commission (No. NCET-13-0048), and the grant from National Natural Science Foundation of China
(No. 11674024 and No. 11574019).
\\

%\bibliographystyle{apsrev}
%\bibliography{reference}

\begin{thebibliography}{36}
\expandafter\ifx\csname natexlab\endcsname\relax\def\natexlab#1{#1}\fi
\expandafter\ifx\csname bibnamefont\endcsname\relax
  \def\bibnamefont#1{#1}\fi
\expandafter\ifx\csname bibfnamefont\endcsname\relax
  \def\bibfnamefont#1{#1}\fi
\expandafter\ifx\csname citenamefont\endcsname\relax
  \def\citenamefont#1{#1}\fi
\expandafter\ifx\csname url\endcsname\relax
  \def\url#1{\texttt{#1}}\fi
\expandafter\ifx\csname urlprefix\endcsname\relax\def\urlprefix{URL }\fi
\providecommand{\bibinfo}[2]{#2}
\providecommand{\eprint}[2][]{\url{#2}}

\bibitem[{\citenamefont{Veselago}(168)}]{Veselago}
\bibinfo{author}{\bibfnamefont{V.~G.} \bibnamefont{Veselago}},
  \bibinfo{journal}{Sov. Phys. Usp.} \textbf{\bibinfo{volume}{10}},
  \bibinfo{pages}{509} (\bibinfo{year}{168}).

\bibitem[{\citenamefont{Ward et~al.}(2005)\citenamefont{Ward, Nelson, and
  Webb}}]{Ward2005}
\bibinfo{author}{\bibfnamefont{D.~W.} \bibnamefont{Ward}},
  \bibinfo{author}{\bibfnamefont{K.~A.} \bibnamefont{Nelson}},
  \bibnamefont{and} \bibinfo{author}{\bibfnamefont{K.~J.} \bibnamefont{Webb}},
  \bibinfo{journal}{New Journal of Physics} \textbf{\bibinfo{volume}{7}},
  \bibinfo{pages}{213} (\bibinfo{year}{2005}), ISSN \bibinfo{issn}{1367-2630},
  \urlprefix\url{http://dx.doi.org/10.1088/1367-2630/7/1/213}.

\bibitem[{\citenamefont{Shelby}(2001)}]{Science292.77-79.Shelby.2001}
\bibinfo{author}{\bibfnamefont{R.~A.} \bibnamefont{Shelby}},
  \bibinfo{journal}{Science} \textbf{\bibinfo{volume}{292}},
  \bibinfo{pages}{77} (\bibinfo{year}{2001}), ISSN \bibinfo{issn}{1095-9203},
  \urlprefix\url{http://dx.doi.org/10.1126/science.1058847}.

\bibitem[{\citenamefont{Smith}(2004)}]{Science305.788-792Smith2004}
\bibinfo{author}{\bibfnamefont{D.~R.} \bibnamefont{Smith}},
  \bibinfo{journal}{Science} \textbf{\bibinfo{volume}{305}},
  \bibinfo{pages}{788} (\bibinfo{year}{2004}), ISSN \bibinfo{issn}{1095-9203},
  \urlprefix\url{http://dx.doi.org/10.1126/science.1096796}.

\bibitem[{\citenamefont{Pendry}(2000)}]{Phys.Rev.Lett.85.3966-3969Pendry2000}
\bibinfo{author}{\bibfnamefont{J.~B.} \bibnamefont{Pendry}},
  \bibinfo{journal}{Phys.Rev.Lett.} \textbf{\bibinfo{volume}{85}},
  \bibinfo{pages}{3966} (\bibinfo{year}{2000}), ISSN \bibinfo{issn}{1079-7114},
  \urlprefix\url{http://dx.doi.org/10.1103/PhysRevLett.85.3966}.

\bibitem[{\citenamefont{Guney and Meyer}(2009)}]{Phys.Rev.A79.Guney2009}
\bibinfo{author}{\bibfnamefont{D.~O.} \bibnamefont{Guney}} \bibnamefont{and}
  \bibinfo{author}{\bibfnamefont{D.~A.} \bibnamefont{Meyer}},
  \bibinfo{journal}{Phys. Rev. A} \textbf{\bibinfo{volume}{79}}
  (\bibinfo{year}{2009}), ISSN \bibinfo{issn}{1094-1622},
  \urlprefix\url{http://dx.doi.org/10.1103/PhysRevA.79.063834}.

\bibitem[{\citenamefont{{Katsnelson} et~al.}(2006)\citenamefont{{Katsnelson},
  {Novoselov}, and {Geim}}}]{Katsnelson2006a}
\bibinfo{author}{\bibfnamefont{M.~I.} \bibnamefont{{Katsnelson}}},
  \bibinfo{author}{\bibfnamefont{K.~S.} \bibnamefont{{Novoselov}}},
  \bibnamefont{and} \bibinfo{author}{\bibfnamefont{A.~K.}
  \bibnamefont{{Geim}}}, \bibinfo{journal}{Nature Physics}
  \textbf{\bibinfo{volume}{2}}, \bibinfo{pages}{620} (\bibinfo{year}{2006}),
  \eprint{cond-mat/0604323}.

\bibitem[{\citenamefont{{Cheianov} et~al.}(2007)\citenamefont{{Cheianov},
  {Fa{\'l}ko}, and {Altshuler}}}]{Cheianov2007a}
\bibinfo{author}{\bibfnamefont{V.~V.} \bibnamefont{{Cheianov}}},
  \bibinfo{author}{\bibfnamefont{V.}~\bibnamefont{{Fa{\'l}ko}}},
  \bibnamefont{and} \bibinfo{author}{\bibfnamefont{B.~L.}
  \bibnamefont{{Altshuler}}}, \bibinfo{journal}{Science}
  \textbf{\bibinfo{volume}{315}}, \bibinfo{pages}{1252} (\bibinfo{year}{2007}),
  \eprint{cond-mat/0703410}.

\bibitem[{\citenamefont{{Xing} et~al.}(2010)\citenamefont{{Xing}, {Wang}, and
  {Sun}}}]{Xing2010}
\bibinfo{author}{\bibfnamefont{Y.}~\bibnamefont{{Xing}}},
  \bibinfo{author}{\bibfnamefont{J.}~\bibnamefont{{Wang}}}, \bibnamefont{and}
  \bibinfo{author}{\bibfnamefont{Q.-F.} \bibnamefont{{Sun}}},
  \bibinfo{journal}{\prb} \textbf{\bibinfo{volume}{81}}, \bibinfo{eid}{165425}
  (\bibinfo{year}{2010}), \eprint{1011.2821}.

\bibitem[{\citenamefont{Fleury and
  Alu}(2014)}]{PhysicalReviewB90.035138.Fleury.2014}
\bibinfo{author}{\bibfnamefont{R.}~\bibnamefont{Fleury}} \bibnamefont{and}
  \bibinfo{author}{\bibfnamefont{A.}~\bibnamefont{Alu}},
  \bibinfo{journal}{Physical Review B} \textbf{\bibinfo{volume}{90}},
  \bibinfo{pages}{035138} (\bibinfo{year}{2014}), ISSN
  \bibinfo{issn}{1550-235X},
  \urlprefix\url{http://dx.doi.org/10.1103/PhysRevB.90.035138}.

\bibitem[{\citenamefont{Pendry et~al.}(2012)\citenamefont{Pendry, Oroszlany,
  Lambert, and Cserti}}]{Pendry2012}
\bibinfo{author}{\bibfnamefont{C.~G.} \bibnamefont{Pendry},
  \bibfnamefont{J.~B.terfalvi}},
  \bibinfo{author}{\bibfnamefont{L.}~\bibnamefont{Oroszlany}},
  \bibinfo{author}{\bibfnamefont{C.~J.} \bibnamefont{Lambert}},
  \bibnamefont{and} \bibinfo{author}{\bibfnamefont{J.}~\bibnamefont{Cserti}},
  \bibinfo{journal}{New Journal of Physics} \textbf{\bibinfo{volume}{14}},
  \bibinfo{pages}{063028} (\bibinfo{year}{2012}), ISSN
  \bibinfo{issn}{1367-2630},
  \urlprefix\url{http://dx.doi.org/10.1088/1367-2630/14/6/063028}.

\bibitem[{\citenamefont{Zhao et~al.}(2015)\citenamefont{Zhao, Wang, Liu, Xu,
  Gu, Xue, and Duan}}]{PhysicalReviewB92.041408.Zhao.2015}
\bibinfo{author}{\bibfnamefont{L.}~\bibnamefont{Zhao}},
  \bibinfo{author}{\bibfnamefont{J.}~\bibnamefont{Wang}},
  \bibinfo{author}{\bibfnamefont{J.}~\bibnamefont{Liu}},
  \bibinfo{author}{\bibfnamefont{Y.}~\bibnamefont{Xu}},
  \bibinfo{author}{\bibfnamefont{B.-L.} \bibnamefont{Gu}},
  \bibinfo{author}{\bibfnamefont{Q.-K.} \bibnamefont{Xue}}, \bibnamefont{and}
  \bibinfo{author}{\bibfnamefont{W.}~\bibnamefont{Duan}},
  \bibinfo{journal}{Physical Review B} \textbf{\bibinfo{volume}{92}},
  \bibinfo{pages}{041408} (\bibinfo{year}{2015}), ISSN
  \bibinfo{issn}{1550-235X},
  \urlprefix\url{http://dx.doi.org/10.1103/PhysRevB.92.041408}.

\bibitem[{\citenamefont{Sessi et~al.}(2016)\citenamefont{Sessi, Ru?mann,
  Bathon, Barla, Kokh, Tereshchenko, Fauth, Mahatha, Valbuena, Godey
  et~al.}}]{PhysicalReviewB94.075137.Sessi.2016}
\bibinfo{author}{\bibfnamefont{P.}~\bibnamefont{Sessi}},
  \bibinfo{author}{\bibfnamefont{P.}~\bibnamefont{Ru?mann}},
  \bibinfo{author}{\bibfnamefont{T.}~\bibnamefont{Bathon}},
  \bibinfo{author}{\bibfnamefont{A.}~\bibnamefont{Barla}},
  \bibinfo{author}{\bibfnamefont{K.~A.} \bibnamefont{Kokh}},
  \bibinfo{author}{\bibfnamefont{O.~E.} \bibnamefont{Tereshchenko}},
  \bibinfo{author}{\bibfnamefont{K.}~\bibnamefont{Fauth}},
  \bibinfo{author}{\bibfnamefont{S.~K.} \bibnamefont{Mahatha}},
  \bibinfo{author}{\bibfnamefont{M.~A.} \bibnamefont{Valbuena}},
  \bibinfo{author}{\bibfnamefont{S.}~\bibnamefont{Godey}},
  \bibnamefont{et~al.}, \bibinfo{journal}{Physical Review B}
  \textbf{\bibinfo{volume}{94}}, \bibinfo{pages}{075137}
  (\bibinfo{year}{2016}), ISSN \bibinfo{issn}{2469-9969},
  \urlprefix\url{http://dx.doi.org/10.1103/PhysRevB.94.075137}.

\bibitem[{\citenamefont{Hasan and
  Kane}(2010)}]{Rev.Mod.Phys.82.3045-3067.Hasan.2010}
\bibinfo{author}{\bibfnamefont{M.~Z.} \bibnamefont{Hasan}} \bibnamefont{and}
  \bibinfo{author}{\bibfnamefont{C.~L.} \bibnamefont{Kane}},
  \bibinfo{journal}{Rev. Mod. Phys.} \textbf{\bibinfo{volume}{82}},
  \bibinfo{pages}{3045} (\bibinfo{year}{2010}), ISSN \bibinfo{issn}{1539-0756},
  \urlprefix\url{http://dx.doi.org/10.1103/RevModPhys.82.3045}.

\bibitem[{\citenamefont{{Qi} and {Zhang}}(2011)}]{Qi2011}
\bibinfo{author}{\bibfnamefont{X.-L.} \bibnamefont{{Qi}}} \bibnamefont{and}
  \bibinfo{author}{\bibfnamefont{S.-C.} \bibnamefont{{Zhang}}},
  \bibinfo{journal}{Reviews of Modern Physics} \textbf{\bibinfo{volume}{83}},
  \bibinfo{pages}{1057} (\bibinfo{year}{2011}), \eprint{1008.2026}.

\bibitem[{\citenamefont{Murakami}(2003)}]{Science301.1348-1351Murakami2003}
\bibinfo{author}{\bibfnamefont{S.}~\bibnamefont{Murakami}},
  \bibinfo{journal}{Science} \textbf{\bibinfo{volume}{301}},
  \bibinfo{pages}{1348} (\bibinfo{year}{2003}), ISSN \bibinfo{issn}{1095-9203},
  \urlprefix\url{http://dx.doi.org/10.1126/science.1087128}.

\bibitem[{\citenamefont{{Bernevig} et~al.}(2006)\citenamefont{{Bernevig},
  {Hughes}, and {Zhang}}}]{Bernevig2006}
\bibinfo{author}{\bibfnamefont{B.~A.} \bibnamefont{{Bernevig}}},
  \bibinfo{author}{\bibfnamefont{T.~L.} \bibnamefont{{Hughes}}},
  \bibnamefont{and} \bibinfo{author}{\bibfnamefont{S.-C.}
  \bibnamefont{{Zhang}}}, \bibinfo{journal}{Science}
  \textbf{\bibinfo{volume}{314}}, \bibinfo{pages}{1757} (\bibinfo{year}{2006}),
  \eprint{cond-mat/0611399}.

\bibitem[{\citenamefont{{K{\"o}nig} et~al.}(2007)\citenamefont{{K{\"o}nig},
  {Wiedmann}, {Br{\"u}ne}, {Roth}, {Buhmann}, {Molenkamp}, {Qi}, and
  {Zhang}}}]{Konig2007}
\bibinfo{author}{\bibfnamefont{M.}~\bibnamefont{{K{\"o}nig}}},
  \bibinfo{author}{\bibfnamefont{S.}~\bibnamefont{{Wiedmann}}},
  \bibinfo{author}{\bibfnamefont{C.}~\bibnamefont{{Br{\"u}ne}}},
  \bibinfo{author}{\bibfnamefont{A.}~\bibnamefont{{Roth}}},
  \bibinfo{author}{\bibfnamefont{H.}~\bibnamefont{{Buhmann}}},
  \bibinfo{author}{\bibfnamefont{L.~W.} \bibnamefont{{Molenkamp}}},
  \bibinfo{author}{\bibfnamefont{X.-L.} \bibnamefont{{Qi}}}, \bibnamefont{and}
  \bibinfo{author}{\bibfnamefont{S.-C.} \bibnamefont{{Zhang}}},
  \bibinfo{journal}{Science} \textbf{\bibinfo{volume}{318}},
  \bibinfo{pages}{766} (\bibinfo{year}{2007}), \eprint{0710.0582}.

\bibitem[{\citenamefont{{Chen} et~al.}(2009)\citenamefont{{Chen}, {Analytis},
  {Chu}, {Liu}, {Mo}, {Qi}, {Zhang}, {Lu}, {Dai}, {Fang} et~al.}}]{Chen2009}
\bibinfo{author}{\bibfnamefont{Y.~L.} \bibnamefont{{Chen}}},
  \bibinfo{author}{\bibfnamefont{J.~G.} \bibnamefont{{Analytis}}},
  \bibinfo{author}{\bibfnamefont{J.-H.} \bibnamefont{{Chu}}},
  \bibinfo{author}{\bibfnamefont{Z.~K.} \bibnamefont{{Liu}}},
  \bibinfo{author}{\bibfnamefont{S.-K.} \bibnamefont{{Mo}}},
  \bibinfo{author}{\bibfnamefont{X.~L.} \bibnamefont{{Qi}}},
  \bibinfo{author}{\bibfnamefont{H.~J.} \bibnamefont{{Zhang}}},
  \bibinfo{author}{\bibfnamefont{D.~H.} \bibnamefont{{Lu}}},
  \bibinfo{author}{\bibfnamefont{X.}~\bibnamefont{{Dai}}},
  \bibinfo{author}{\bibfnamefont{Z.}~\bibnamefont{{Fang}}},
  \bibnamefont{et~al.}, \bibinfo{journal}{Science}
  \textbf{\bibinfo{volume}{325}}, \bibinfo{pages}{178} (\bibinfo{year}{2009}).

\bibitem[{\citenamefont{{Zhang} et~al.}(2009)\citenamefont{{Zhang}, {Liu},
  {Qi}, {Dai}, {Fang}, and {Zhang}}}]{Zhang2009}
\bibinfo{author}{\bibfnamefont{H.}~\bibnamefont{{Zhang}}},
  \bibinfo{author}{\bibfnamefont{C.-X.} \bibnamefont{{Liu}}},
  \bibinfo{author}{\bibfnamefont{X.-L.} \bibnamefont{{Qi}}},
  \bibinfo{author}{\bibfnamefont{X.}~\bibnamefont{{Dai}}},
  \bibinfo{author}{\bibfnamefont{Z.}~\bibnamefont{{Fang}}}, \bibnamefont{and}
  \bibinfo{author}{\bibfnamefont{S.-C.} \bibnamefont{{Zhang}}},
  \bibinfo{journal}{Nature Physics} \textbf{\bibinfo{volume}{5}},
  \bibinfo{pages}{438} (\bibinfo{year}{2009}).

\bibitem[{\citenamefont{Yu et~al.}(2010)\citenamefont{Yu, Zhang, Zhang, Zhang,
  Dai, and Fang}}]{Science329.61-64.Yu.2010}
\bibinfo{author}{\bibfnamefont{R.}~\bibnamefont{Yu}},
  \bibinfo{author}{\bibfnamefont{W.}~\bibnamefont{Zhang}},
  \bibinfo{author}{\bibfnamefont{H.-J.} \bibnamefont{Zhang}},
  \bibinfo{author}{\bibfnamefont{S.-C.} \bibnamefont{Zhang}},
  \bibinfo{author}{\bibfnamefont{X.}~\bibnamefont{Dai}}, \bibnamefont{and}
  \bibinfo{author}{\bibfnamefont{Z.}~\bibnamefont{Fang}},
  \bibinfo{journal}{Science} \textbf{\bibinfo{volume}{329}},
  \bibinfo{pages}{61} (\bibinfo{year}{2010}),
  \urlprefix\url{http://dx.doi.org/10.1126/science.1187485}.

\bibitem[{\citenamefont{Chang et~al.}(2013)\citenamefont{Chang, Zhang, Feng,
  Shen, Zhang, Guo, Li, Ou, Wei, Wang et~al.}}]{Science340.167-170.Chang.2013}
\bibinfo{author}{\bibfnamefont{C.-Z.} \bibnamefont{Chang}},
  \bibinfo{author}{\bibfnamefont{J.}~\bibnamefont{Zhang}},
  \bibinfo{author}{\bibfnamefont{X.}~\bibnamefont{Feng}},
  \bibinfo{author}{\bibfnamefont{J.}~\bibnamefont{Shen}},
  \bibinfo{author}{\bibfnamefont{Z.}~\bibnamefont{Zhang}},
  \bibinfo{author}{\bibfnamefont{M.}~\bibnamefont{Guo}},
  \bibinfo{author}{\bibfnamefont{K.}~\bibnamefont{Li}},
  \bibinfo{author}{\bibfnamefont{Y.}~\bibnamefont{Ou}},
  \bibinfo{author}{\bibfnamefont{P.}~\bibnamefont{Wei}},
  \bibinfo{author}{\bibfnamefont{L.-L.} \bibnamefont{Wang}},
  \bibnamefont{et~al.}, \bibinfo{journal}{Science}
  \textbf{\bibinfo{volume}{340}}, \bibinfo{pages}{167} (\bibinfo{year}{2013}),
  \urlprefix\url{http://dx.doi.org/10.1126/science.1234414}.

\bibitem[{\citenamefont{Zhang et~al.}(2010)\citenamefont{Zhang, Yu, Zhang, Dai,
  and Fang}}]{Zhang2010}
\bibinfo{author}{\bibfnamefont{W.}~\bibnamefont{Zhang}},
  \bibinfo{author}{\bibfnamefont{R.}~\bibnamefont{Yu}},
  \bibinfo{author}{\bibfnamefont{H.-J.} \bibnamefont{Zhang}},
  \bibinfo{author}{\bibfnamefont{X.}~\bibnamefont{Dai}}, \bibnamefont{and}
  \bibinfo{author}{\bibfnamefont{Z.}~\bibnamefont{Fang}}, \bibinfo{journal}{New
  Journal of Physics} \textbf{\bibinfo{volume}{12}}, \bibinfo{pages}{065013}
  (\bibinfo{year}{2010}), ISSN \bibinfo{issn}{1367-2630},
  \urlprefix\url{http://dx.doi.org/10.1088/1367-2630/12/6/065013}.

\bibitem[{\citenamefont{Shan et~al.}(2010)\citenamefont{Shan, Lu, and
  Shen}}]{Shan2010}
\bibinfo{author}{\bibfnamefont{W.-Y.} \bibnamefont{Shan}},
  \bibinfo{author}{\bibfnamefont{H.-Z.} \bibnamefont{Lu}}, \bibnamefont{and}
  \bibinfo{author}{\bibfnamefont{S.-Q.} \bibnamefont{Shen}},
  \bibinfo{journal}{New Journal of Physics} \textbf{\bibinfo{volume}{12}},
  \bibinfo{pages}{043048} (\bibinfo{year}{2010}), ISSN
  \bibinfo{issn}{1367-2630},
  \urlprefix\url{http://dx.doi.org/10.1088/1367-2630/12/4/043048}.

\bibitem[{\citenamefont{{Liu} et~al.}(2010)\citenamefont{{Liu}, {Qi}, {Zhang},
  {Dai}, {Fang}, and {Zhang}}}]{Liu2010}
\bibinfo{author}{\bibfnamefont{C.-X.} \bibnamefont{{Liu}}},
  \bibinfo{author}{\bibfnamefont{X.-L.} \bibnamefont{{Qi}}},
  \bibinfo{author}{\bibfnamefont{H.-J.} \bibnamefont{{Zhang}}},
  \bibinfo{author}{\bibfnamefont{X.}~\bibnamefont{{Dai}}},
  \bibinfo{author}{\bibfnamefont{Z.}~\bibnamefont{{Fang}}}, \bibnamefont{and}
  \bibinfo{author}{\bibfnamefont{S.-C.} \bibnamefont{{Zhang}}},
  \bibinfo{journal}{\prb} \textbf{\bibinfo{volume}{82}}, \bibinfo{eid}{045122}
  (\bibinfo{year}{2010}), \eprint{1005.1682}.

\bibitem[{\citenamefont{{Zhang}
  et~al.}(2014{\natexlab{a}})\citenamefont{{Zhang}, {Zhuang}, {Xing}, {Li},
  {Wang}, and {Guo}}}]{Zhang2014a}
\bibinfo{author}{\bibfnamefont{L.}~\bibnamefont{{Zhang}}},
  \bibinfo{author}{\bibfnamefont{J.}~\bibnamefont{{Zhuang}}},
  \bibinfo{author}{\bibfnamefont{Y.}~\bibnamefont{{Xing}}},
  \bibinfo{author}{\bibfnamefont{J.}~\bibnamefont{{Li}}},
  \bibinfo{author}{\bibfnamefont{J.}~\bibnamefont{{Wang}}}, \bibnamefont{and}
  \bibinfo{author}{\bibfnamefont{H.}~\bibnamefont{{Guo}}},
  \bibinfo{journal}{\prb} \textbf{\bibinfo{volume}{89}}, \bibinfo{eid}{245107}
  (\bibinfo{year}{2014}{\natexlab{a}}).

\bibitem[{\citenamefont{{Chen} et~al.}(2012)\citenamefont{{Chen}, {Wang}, and
  {Sun}}}]{Chen2012}
\bibinfo{author}{\bibfnamefont{J.-C.} \bibnamefont{{Chen}}},
  \bibinfo{author}{\bibfnamefont{J.}~\bibnamefont{{Wang}}}, \bibnamefont{and}
  \bibinfo{author}{\bibfnamefont{Q.-F.} \bibnamefont{{Sun}}},
  \bibinfo{journal}{\prb} \textbf{\bibinfo{volume}{85}}, \bibinfo{eid}{125401}
  (\bibinfo{year}{2012}).

\bibitem[{\citenamefont{{Zhang}
  et~al.}(2014{\natexlab{b}})\citenamefont{{Zhang}, {Jiang}, {Xie}, and
  {Sun}}}]{Zhang2014}
\bibinfo{author}{\bibfnamefont{S.-f.} \bibnamefont{{Zhang}}},
  \bibinfo{author}{\bibfnamefont{H.}~\bibnamefont{{Jiang}}},
  \bibinfo{author}{\bibfnamefont{X.~C.} \bibnamefont{{Xie}}}, \bibnamefont{and}
  \bibinfo{author}{\bibfnamefont{Q.-f.} \bibnamefont{{Sun}}},
  \bibinfo{journal}{\prb} \textbf{\bibinfo{volume}{89}}, \bibinfo{eid}{155419}
  (\bibinfo{year}{2014}{\natexlab{b}}).

\bibitem[{\citenamefont{{Jauho} et~al.}(1994)\citenamefont{{Jauho}, {Wingreen},
  and {Meir}}}]{Jauho1994}
\bibinfo{author}{\bibfnamefont{A.-P.} \bibnamefont{{Jauho}}},
  \bibinfo{author}{\bibfnamefont{N.~S.} \bibnamefont{{Wingreen}}},
  \bibnamefont{and} \bibinfo{author}{\bibfnamefont{Y.}~\bibnamefont{{Meir}}},
  \bibinfo{journal}{\prb} \textbf{\bibinfo{volume}{50}}, \bibinfo{pages}{5528}
  (\bibinfo{year}{1994}), \eprint{cond-mat/9404027}.

\bibitem[{\citenamefont{M.~P. Lopez~Sancho and
  Rubio}(1984)}]{J.Phys.F:Met.Phys.1984}
\bibinfo{author}{\bibfnamefont{J.~M. L.~S.} \bibnamefont{M.~P. Lopez~Sancho}}
  \bibnamefont{and} \bibinfo{author}{\bibfnamefont{J.}~\bibnamefont{Rubio}},
  \bibinfo{journal}{J.Phys.F: Met.Phys.} \textbf{\bibinfo{volume}{14}},
  \bibinfo{pages}{1205} (\bibinfo{year}{1984}).

\bibitem[{\citenamefont{M.~P. Lopez~Sancho and
  Rubio}(1985)}]{J.Phys.F:Met.Phys.1985}
\bibinfo{author}{\bibfnamefont{J.~M. L.~S.} \bibnamefont{M.~P. Lopez~Sancho}}
  \bibnamefont{and} \bibinfo{author}{\bibfnamefont{J.}~\bibnamefont{Rubio}},
  \bibinfo{journal}{J.Phys.F: Met.Phys.} \textbf{\bibinfo{volume}{15}},
  \bibinfo{pages}{851} (\bibinfo{year}{1985}).

\bibitem[{\citenamefont{{Lee} and
  {Joannopoulos}}(1981{\natexlab{a}})}]{Lee1981}
\bibinfo{author}{\bibfnamefont{D.~H.} \bibnamefont{{Lee}}} \bibnamefont{and}
  \bibinfo{author}{\bibfnamefont{J.~D.} \bibnamefont{{Joannopoulos}}},
  \bibinfo{journal}{\prb} \textbf{\bibinfo{volume}{23}}, \bibinfo{pages}{4988}
  (\bibinfo{year}{1981}{\natexlab{a}}).

\bibitem[{\citenamefont{{Lee} and
  {Joannopoulos}}(1981{\natexlab{b}})}]{Lee1981a}
\bibinfo{author}{\bibfnamefont{D.~H.} \bibnamefont{{Lee}}} \bibnamefont{and}
  \bibinfo{author}{\bibfnamefont{J.~D.} \bibnamefont{{Joannopoulos}}},
  \bibinfo{journal}{\prb} \textbf{\bibinfo{volume}{23}}, \bibinfo{pages}{4997}
  (\bibinfo{year}{1981}{\natexlab{b}}).

\bibitem[{\citenamefont{Xing et~al.}(2011)\citenamefont{Xing, Zhang, and
  Wang}}]{Xing2011a}
\bibinfo{author}{\bibfnamefont{Y.}~\bibnamefont{Xing}},
  \bibinfo{author}{\bibfnamefont{L.}~\bibnamefont{Zhang}}, \bibnamefont{and}
  \bibinfo{author}{\bibfnamefont{J.}~\bibnamefont{Wang}},
  \bibinfo{journal}{Physical Review B} \textbf{\bibinfo{volume}{84}},
  \bibinfo{pages}{035110} (\bibinfo{year}{2011}), ISSN
  \bibinfo{issn}{1550-235X},
  \urlprefix\url{http://dx.doi.org/10.1103/PhysRevB.84.035110}.

\bibitem[{\citenamefont{Chen et~al.}(2012)\citenamefont{Chen, Liu, Lin, Zhang,
  and Jiang}}]{Chen2012a}
\bibinfo{author}{\bibfnamefont{L.}~\bibnamefont{Chen}},
  \bibinfo{author}{\bibfnamefont{Q.}~\bibnamefont{Liu}},
  \bibinfo{author}{\bibfnamefont{X.}~\bibnamefont{Lin}},
  \bibinfo{author}{\bibfnamefont{X.}~\bibnamefont{Zhang}}, \bibnamefont{and}
  \bibinfo{author}{\bibfnamefont{X.}~\bibnamefont{Jiang}},
  \bibinfo{journal}{New Journal of Physics} \textbf{\bibinfo{volume}{14}},
  \bibinfo{pages}{043028} (\bibinfo{year}{2012}), ISSN
  \bibinfo{issn}{1367-2630},
  \urlprefix\url{http://dx.doi.org/10.1088/1367-2630/14/4/043028}.

\bibitem[{\citenamefont{Xiao et~al.}(2010)\citenamefont{Xiao, Chang, and
  Niu}}]{Rev.Mod.Phys.82.1959-2007.Xiao.2010}
\bibinfo{author}{\bibfnamefont{D.}~\bibnamefont{Xiao}},
  \bibinfo{author}{\bibfnamefont{M.-C.} \bibnamefont{Chang}}, \bibnamefont{and}
  \bibinfo{author}{\bibfnamefont{Q.}~\bibnamefont{Niu}}, \bibinfo{journal}{Rev.
  Mod. Phys.} \textbf{\bibinfo{volume}{82}}, \bibinfo{pages}{1959}
  (\bibinfo{year}{2010}), ISSN \bibinfo{issn}{1539-0756},
  \urlprefix\url{http://dx.doi.org/10.1103/RevModPhys.82.1959}.

\end{thebibliography}

\end{document}